\documentclass[12pt]{article}
\usepackage{amsfonts}
\usepackage{amssymb}
\usepackage{amsbsy}
\usepackage{graphicx}    
\usepackage{rotating}    
\usepackage{epsfig}
\usepackage{color}

\input epsf.sty

\newlength{\TZ}
\newcommand{\BEQ}{\begin{equation}}     
\newcommand{\BEA}{\begin{eqnarray}}
\newcommand{\BD}{\begin{displaymath}}
\newcommand{\EEQ}{\end{equation}}       
\newcommand{\EEA}{\end{eqnarray}}
\newcommand{\ED}{\end{displaymath}}
\newcommand{\bb}{\begin{eqnarray}}
\newcommand{\ee}{\end{eqnarray}}

\newcommand{\D}{{\rm d}}                
\newcommand{\II}{{\rm i}}               
 
\newcommand{\demi}{\frac{1}{2}}         
\newcommand{\wit}[1]{\widetilde{#1}}    
\newcommand{\wht}[1]{\widehat{#1}}      
\newcommand{\lap}[1]{\overline{#1}}     

\renewcommand{\vec}[1]{\boldsymbol{#1}} 

\newcommand{\appsection}[2]{\setcounter{equation}{0}\setcounter{subsection}{0}
\section*{Appendix #1. #2}
\renewcommand{\theequation}{#1.\arabic{equation}}
              \renewcommand{\thesection}{#1} }

\catcode`\@=11
\def\numberbysection{\@addtoreset{equation}{section}
        \def\theequation{\thesection.\arabic{equation}}}
\numberbysection

\begin{document}

\begin{titlepage}

\vskip 1.5 cm
\begin{center}
{\Large \bf Spherical model of growing interfaces}
\end{center}

\vskip 2.0 cm
\centerline{{\bf Malte Henkel}$^a$\footnote{e-mail: malte.henkel@univ-lorraine.fr} 
and {\bf Xavier Durang}$^b$\footnote{e-mail: xdurang@kias.re.kr}}
\vskip 0.5 cm
\begin{center}
$^a$Groupe de Physique Statistique,
D\'epartement de Physique de la Mati\`ere et des Mat\'eriaux,
Institut Jean Lamour (CNRS UMR 7198), Universit\'e de Lorraine Nancy, 
B.P. 70239, \\ F -- 54506 Vand{\oe}uvre l\`es Nancy Cedex, France\\
$^b$School of Physics, Korea Institute for Advanced Study, Seoul 130-722, Korea\\~\\
\end{center}

\begin{abstract}
Building on an analogy between the ageing behaviour of magnetic systems and growing interfaces, the Arcetri model, 
a new exactly solvable model for growing interfaces is introduced, 
which shares many properties with the kinetic spherical model. 
The long-time behaviour of the interface width and of the two-time correlators and responses is analysed. 
For all dimensions $d\ne 2$, universal characteristics distinguish the Arcetri model from the Edwards-Wilkinson model, 
although for $d>2$ all stationary and non-equilibrium exponents are the same.  
For $d=1$ dimensions, the Arcetri model is equivalent to the $p=2$ spherical spin glass. 
For $2<d<4$ dimensions, its relaxation properties are related to the ones of a particle-reaction model, 
namely a bosonic variant of the diffusive pair-contact process. The global persistence exponent is also derived. 
\end{abstract}

\vfill

\end{titlepage}

\setcounter{footnote}{0} 

\section{Introduction}

The physics of the growth of interfaces is a paradigmatic example of 
the emergence of non-equilibrium cooperative phenomena 
\cite{Bara95,Halp95,Krug97,Krie10,Corw12,Wio13,Taeu14}.\footnote{Applications include 
settings as distinct as deposition of atoms on a surface, 
solidification, flame propagation, population dynamics, crack propagation, 
chemical reaction fronts or the growth of cell colonies.} 
The morphological evolution of 
the growing interface is characterised by its fractal properties as well
as several growth and roughness exponents. More recently, aspects of the 
time-dependent evolution of the interface have been investigated
and have been found to be quite analogous to phenomena encountered in the 
physical ageing in glassy and non-glassy systems \cite{Cugl03,Henk10}. 
It has been understood that growing interfaces can be cast into 
distinct universality classes, including those associated to the
Edwards-Wilkinson ({\sc ew}) equation \cite{Edwa82}, where the 
relaxation is dominated by the linear interface tension, 
and the Kardar-Parisi-Zhang ({\sc kpz}) equation \cite{Kard86}, 
which includes as well a non-linear effect of the local slope. 

The exponents of growing interfaces are defined as follows. 
The basic quantity is the time-space-dependent local height $h(t,\vec{r})$ and one is
mainly interested in the fluctuations around the spatially averaged height 
$\overline{h}(t) := L^{-d} \sum_{\vec{r}\in\Lambda} h(t,\vec{r})$; 
for notational convenience defined here on a hyper-cubic lattice $\Lambda\subset\mathbb{Z}^d$ with $L^d$ sites 
and the sum runs over the lattice sites. A central quantity is the interface width
\BEQ \label{1.1}
w^2(t;L) := \frac{1}{L^d} \sum_{\vec{r}\in\Lambda} \left\langle \left( h(t,\vec{r})-\overline{h}(t) \right) \right\rangle^2 
= L^{2\alpha} f_w\left(t L^{-z}\right) 
\sim \left\{ \begin{array}{ll} t^{2\beta} & \mbox{\rm ~;~~ if $tL^{-z}\ll 1$} \\
                              L^{2\alpha} & \mbox{\rm ~;~~ if $tL^{-z}\gg 1$} \end{array} \right.
\EEQ
\textcolor{black}{and we recall} 
the generically expected Family-Viscek scaling form \cite{Fami85}, 
where $\alpha$ is the roughness exponent, $\beta$ the growth exponent and
$z=\alpha/\beta$ the dynamical exponent. Herein, $\langle .\rangle$ denotes an average over many independent samples 
\textcolor{black}{(under the same thermodynamic conditions)}. 
Throughout this work, we shall consider the $L\to\infty$ limit. The initial state will always be a flat, uncorrelated substrate. 
Relaxational properties of the interface can be characterised by the two-time 
correlations and (linear) responses
\BEA
C(t,s;\vec{r}) &:=& \left\langle \left(h(t,\vec{r}) - \left\langle \overline{h}(t)\right\rangle \right)
\left(h(s,\vec{0}) - \left\langle \overline{h}(s)\right\rangle \right) \right\rangle 
\:=\: s^{-b} F_C\left( \frac{t}{s}; \frac{\vec{r}}{s^{1/z}} \right)  \label{1.2} \\
R(t,s;\vec{r}) &:=& \left. \frac{ \delta \left\langle h(t,\vec{r}) - \overline{h}(t)\right\rangle}{\delta j(s,\vec{0})}\right|_{j=0} 
\:=\: \left\langle h(t,\vec{r}) \wit{h}(s,\vec{0}) \right\rangle \:=\: s^{-1-a} F_R\left( \frac{t}{s}; \frac{\vec{r}}{s^{1/z}} \right)
\label{1.3}
\EEA
where spatial translation-invariance has been implicitly admitted and $j$ is an external field conjugate to $h$. 
Furthermore, in the context of Janssen-de Dominicis theory, $\wit{h}$ is the conjugate response field to $h$ \cite{Jans89,Taeu14}. 
In the long-time limit, where both $t,s\gg \tau_{\rm micro}$ and
$t-s\gg \tau_{\rm micro}$ ($\tau_{\rm micro}$ is a microscopic reference time), 
generalised Family-Viscek scaling forms were assumed \cite{Kall99,Roet06,Bust07,Igua09,Henk12} which are analogous to the scaling forms of 
`simple ageing' in other non-equilibrium systems \cite{Cugl03,Cugl09,Henk10}. 
The exponent $b=-2\beta$ \cite{Kall99,Daqu11}, but the relationship of $a$ 
to other exponents seems to depend on the universality class \cite{Henk12}. 
Finally, the autocorrelation exponent $\lambda_C$ and the autoresponse exponent $\lambda_R$ are defined from the asymptotics
$F_{C,R}(y,\vec{0}) \sim y^{-\lambda_{C,R}/z}$ as $y\to\infty$. There is a bound $\lambda_C\geq (d+zb)/2$, see appendix~B. 
Field-theoretical considerations in ageing simple magnets with a non-conserved model-A dynamics and with disordered initial conditions
strongly indicate that the non-equilibrium exponents $\lambda_C,\lambda_R$ should be independent of those describing the stationary state
\cite{Jans89,Cala05,Taeu14}. This is different, however, for the {\sc kpz} universality class, where for dimensions $d<2$ it was shown that
$\lambda_C=d$, to all orders in perturbation-theory \cite{Krec97}. However, for $d\geq 2$ there exists a strong-coupling fixed point of the 
{\sc kpz} equation which cannot be reached by a perturbative analysis, see e.g. \cite{Wies98,Halp13,Taeu14}; 
but which can be studied through non-perturbative renormalisation-group techniques \cite{Cane11,Klos12}. 

\begin{sidewaystable}
\caption[tab1]{Exponents of growing interfaces in four universality classes, 
namely Kardar-Parisi-Zhang ({\sc kpz}), (positive) quenched KPZ ({\sc qkpz}),
Edwards-Wilkinson ({\sc ew}) and Arcetri (for both $T=T_c$ and $T<T_c$) are 
represented. In each group, labelled by the universality class, 
first are indicated available exact results$^{\dag}$ and/or recent simulational estimates.$^{\dag}$ 
Then follow experimental results, where the system's dimension $d$ and the nature of the interface are indicated. 
The numbers in bracket give the error in the last digit(s). 
\\ 
{\footnotesize $^*$The scaling relation $\alpha=z\beta$, resp. $\alpha+z=2$ for the KPZ class, was used to 
complete the entries as much as possible, 
with an error estimated from the relative errors of the exponent(s) given in the source.}\\ \label{tab1}} 
\begin{tabular}{||l|cccccccc|l||} \hline\hline
model       & ~$d$~ & ~$z$~        & ~$\beta$~      & ~$\alpha$~ & ~$a$~     & ~$b$~       & ~$\lambda_C$~ & ~$\lambda_R$~ & Ref. \\ \hline\hline
KPZ         & $1$   & $3/2$        & $1/3$          & $1/2$      & $-1/3$    & $-2/3$      & $1$           & $1$       & \cite{Kard86,Krec97,Henk12} \\
            & $2$   & $1.61(2)^*$~ & $0.2415(15)$   & $0.393(4)$ & $0.30(1)$ & $-0.483(3)$ & $1.97(3)$     & $2.04(3)$ & \cite{Odor14} \\
            & $2$   & $1.61(2)^*$~ & $0.241(1)$~~~  & $0.393(3)$ &           & $-0.483$~~~ & $1.91(6)$     &           & \cite{Halp14} \\ 
            & $2$   & $1.61(5)$~~  & $0.244(2)$~~~  & $0.369(8)$ &           &             &               &           & \cite{Rodr15} \\
            & $2$   & $1.627(4)^*$ & $0.229(6)^*$~~ & $0.373(3)$ &           &             &               &           & \cite{Klos12} \\ \hline
{\footnotesize Ag electrodeposition}$^{\dag}$
            & $1$   &              & ${\small\approx 1/3}$~~ & ${\small\approx 1/2}$~~ & & &               &           & \cite{Schi99} \\
{\footnotesize slow paper combustion}$^{\dag}$ 
            & $1$   & $1.44(12)^*$ & $0.32(4)$~~~~  & $0.49(4)$~~ &          &             &               &           & \cite{Maun97} \\
{\footnotesize liquid crystal}$^{\dag}$ 
            & $1$   & $1.34(14)^*$ & $0.32(2)$~~~~  & $0.43(6)$~~ & & ${\small\approx -2/3}$ & ${\small\approx 1}$~ &  & \cite{Take12} \\ 
{\footnotesize liquid crystal}$^{\ddag}$ 
            & $1$   & $1.44(10)^*$ & $0.334(3)$~~~  & $0.48(5)$~~ & & ${\small\approx -2/3}$ & $0${\rm ?}  &           & \cite{Take12} \\ 
{\footnotesize cell colony growth}$^{\dag \ddag}$
            & $1$  & $1.56(10)$~   & $0.32(4)$~~~~~ & $0.50(5)$~~ &          &             &               &           & \cite{Huer12,Huer14} \\
{\footnotesize (almost) isotrope collo\"{\i}ds}$^{\ddag}$ 
            & $1$   &              & $0.37(4)$~~~~~ & $0.51(5)$~~ &          &             &               &           & \cite{Yunk13} \\
{\footnotesize autocatalytic reaction front}$^{\dag}$ 
            & $1$   & $1.45(11)^*$ & $0.34(4)$~~~~~ & $0.50(4)$~~ &          &             &               &           & \cite{Atis14} \\
{\footnotesize CdTe/Si(100) film}$^{\dag}$ 
            & $2$   & $1.61(5)$~~  & $0.24(4)$~~~~~ & $0.39(5)^*$~ &         &             &               &           & \cite{Alme13} \\ \hline\hline
QKPZ        & $1$   &         & $\approx 0.9$~~~~~  & $0.63261$~  &          &             &               &           & \cite{Tang92,Snep92} \\ \hline 
{\footnotesize cell colony growth (disordered)}$^{\dag}$ 
            & $1$   & $0.84(5)$~~  & $0.75(5)$~~~~~ & $0.63(4)$~~ &          &             &               &           & \cite{Huer14} \\  
{\footnotesize autocatalytic reaction front}$^{\dag}$ 
            & $1$   &              & $0.61(5)$~~~~~ & $0.66(4)$~~ &          &             &               &           & \cite{Atis14} \\
{\footnotesize strongly ellipsoid collo\"{\i}ds}$^{\ddag}$ 
            & $1$   &              & $0.68(5)$~~~~~ & $0.61(2)$~~ &          &             &               &           & \cite{Yunk13} \\ \hline\hline
EW          & $<2$  & $2$          & $(2-d)/4$      & $(2-d)/2$   & $d/2-1$  & $d/2-1$     & $d$           & $d$       &  \\
            & $2$   & $2$         & $0$(log)$^{\#}$ & $0$(log)$^{\#}$ & $0$  & $0$         & $2$           & $2$       & \cite{Edwa82,Roet06} \\ 
            & $>2$  & $2$          & $0$            & $0$         & $d/2-1$  & $d/2-1$     & $d$           & $d$       &  \\ \hline
{\footnotesize sedimentation/electrodispersion}
            & $2$   &             & $0$(log)$^{\#}$ & $0$(log)$^{\#}$    &   &             &               &           & \cite{Salv96} \\
\hline\hline\hline
{\bf Arcetri} \hfill $T=T_c$\, 
            & $<2$  & $2$          & $(2-d)/4$      & $(2-d)/2$   & $d/2-1$  & $d/2-1$     & $3d/2-1$      & $3d/2-1$  & \\ 
            & $2$  & $2$          & $0$(log)$^{\#}$ & 
                                                                  & $0$      & $0$         & $2$           & $2$       & \\ 
            & $>2$  & $2$          & $0$            & $0$         & $d/2-1$  & $d/2-1$     & $d$           & $d$       & \\[0.19cm]
\hfill $T<T_c$\, & $d$ & $2$       & $1/2$          & $1$         & $d/2-1$  & $-1$~~      & $d/2-1$       & $d/2-1$   & \\ \hline\hline
\end{tabular} 
~\\
{\footnotesize$^{\dag}$flat interface.~ $^{\ddag}$circular interface. 
$^{\#}$For $d=2$, one has $w(t;L)\sim \sqrt{\ln t\,}\: f_w\left( \ln L/\ln t\right)$.} 
\end{sidewaystable}

Known estimates of all these exponents, for several universality classes, are listed in table~\ref{tab1}. 
In most of the quoted experiments, two-time autocorrelators such as $C(t,0;\vec{0})$ 
were also used to extract $\beta$, these approaches rely on the relation $b=-2\beta$.\footnote{However, this relation needs no longer
to hold true when $d>d^*$.} The study of two-time global quantities, also of interest for experiments, 
gives a different access to the exponents, especially $\beta$ \cite{Chou09,Assi14}. 
For the experiments reported in \cite{Atis14,Huer12,Take12},
merely an average of the values quoted in the sources is included in table~\ref{tab1}. 
While the theoretical analysis, at least for the exponents $\alpha,\beta,z$
in the {\sc ew} and {\sc kpz} classes, has been achieved long ago \cite{Edwa82,Kard86}, 
recent advances in experimental techniques have permitted
to obtain precise and reliable estimates of these exponents, independently and in several distinct 
systems,\footnote{In \cite{Huer12}, \textcolor{black}{colonies} of both benign and malign \textcolor{black}{cells}, 
for both flat and circular interfaces were measured.} 
and in good agreement with each other and with the 
theoretical predictions.\footnote{See \cite{Alve14,Kim14,Rodr15} for recent estimates, 
of the exponents $\alpha,\beta$ and also of universal amplitude ratios, in the {\sc kpz} universality class, up to $d=11$.} 
Notably, the great variety
of systems for which the exponents $z,\beta,\alpha$ are consistent with the {\sc kpz}-equation, clearly attest its universality, 
\textcolor{black}{at least for $1D$ systems}. 
For the $1D$ (positive) {\bf q}uenched {\sc kpz} class, a mapping onto $1D$ directed percolation was proposed \cite{Tang92}. This produces 
$\alpha=\nu_{\perp}/\nu_{\|}\simeq 0.63261$ which apparently fits the experimental and simulational data well 
(using the very precisely known values of
the exponents $\nu_{\|,\perp}$ of directed percolation, see e.g. \cite{Henk09}). However, 
other predictions following from this mapping, such as $z=1$ \cite{Tang92,Bara95}, 
do not seem to be generically reproduced by the presently available evidence. 
In several of these experiments, universal amplitude ratios have been measured as well \cite{Alme13,Take11,Take12,Yunk13}. 
For further details on procedures and the extraction of the exponents, we refer to the quoted sources.  
For a recent thorough review on experimental results, see \cite{Take14}. 

In $d=1$ dimension, several further notable advances have been achieved. 
These concern a remarkable exact solution of the {\sc kpz} equation
and the spectacular relationship of the probability distribution ${\cal P}(h)$ of the 
fluctuation $h-\overline{h}$ with the extremal value statistics of the largest eigenvalue of random matrices, 
see \cite{Sasa10,Cala11,Cala14,Imam12,Halp12,Halp13},
and its successful experimental confirmation \cite{Take11,Take12,Take14,Yunk13,Halp14,Halp14b}. 
Still, this progress seems to rely on specific properties of the one-dimensional case.
One may ask, if there might exist alternative routes to a further conceptual understanding, 
less dependent on the particular mathematical circumstances
which render the $1D$ {\sc kpz} equation analytically treatable. 
Here, an analogy with magnetic systems at their critical point might be helpful. 
Some elements of such an analogy are sketched in table~\ref{tab2}. 
\begin{table}[tb]
\caption[tab2]{Analogies between the critical dynamics in magnets and growing interfaces. 
Several observables derived from the central physical quantity, the order parameter/height, respectively, 
are shown, where $h$ or $j$, respectively, are conjugate to the order parameter/height. 
The average $\langle .\rangle_c$ denotes a connected correlator.  
Some models, with the equilibrium hamiltonian for magnets, are defined through their kinetic equations, 
where $\eta$ is a standard white noise, $\Delta$ the spatial laplacian and $D,g,\nu,\mu$ are constants. 
The existence of known exact solutions in the Ising and {\sc kpz} models is indicated. \\
\label{tab2}} 
\begin{tabular}{|ll|ll|} \hline
\multicolumn{2}{|c|}{magnets} & \multicolumn{2}{c|}{interfaces} \\ \hline
{\small order parameter} & $\phi(t,\vec{r})$  & {\small height} & $h(t,\vec{r})$ \\
{\small variance}        & $\left\langle ( \phi(t,\vec{r})-\left\langle \phi(t,\vec{r})\right\rangle )^2\right\rangle \sim t^{-2\beta/(\nu z)}$ &
{\small width} & $w^2(t) = \langle (h(t,\vec{r})-\overline{h}(t))^2\rangle \sim t^{2\beta}$ \\
{\small autocorrelator}  & $C(t,s) = \left\langle \phi(t,\vec{r}) \phi(s,\vec{r})\right\rangle_c$ & 
{\small autocorrelator}  & $C(t,s) = \left\langle h(t,\vec{r}) h(s,\vec{r}) \right\rangle_c$ \\
{\small autoresponse}    & $R(t,s) = \left. \delta \langle \phi(t,\vec{r})\rangle/\delta h(s,\vec{r})\right|_{h=0}$ & 
{\small autoresponse}    & $R(t,s) = \left. \delta \langle h(t,\vec{r})\rangle/\delta j(s,\vec{r})\right|_{j=0}$ \\ \hline
\multicolumn{4}{|l|}{{\bf Models}:} \\[0.18cm] 
gaussian field   & ${\cal H}[\phi] = -\demi \int \!\D\vec{r}\, (\nabla\phi)^2$ & {\sc ew} & \\
                 & $\partial_t\phi = D \Delta\phi +\eta$                       &          & $\partial_t h = \nu \Delta h +\eta$ \\[0.15cm]
Ising model      & ${\cal H}[\phi] = -\demi \int \!\D\vec{r}\, [(\nabla\phi)^2 +\frac{g}{2} \phi^4]$ & {\sc kpz} & \\
                 & $\partial_t\phi = D (\Delta\phi+ g\phi^3) +\eta$  & & $\partial_t h = \nu\Delta h +\frac{\mu}{2} (\nabla h)^2 +\eta$ \\ \hline
\multicolumn{4}{|l|}{{\bf Exact solutions} for the Ising and {\sc kpz} universality classes:} \\[0.18cm]
\multicolumn{2}{|l|}{equilibrium $d=2$ \hfill \cite{Onsa44,Zamo89,Warn94}}&  & \\
\multicolumn{2}{|l|}{relaxation~~  $d=1$ \hfill \cite{Glau63}}  &
\multicolumn{2}{l|}{relaxation $d=1$  \hfill \cite{Sasa10,Cala11,Imam12}} \\ \hline
\end{tabular} 
\end{table}
Indeed, several observables derived from the main quantity, namely the order parameter $\phi(t,\vec{r})$ 
or the interface height $h(t,\vec{r})$, respectively, show an analogous dynamical behaviour. In particular, 
the exponent $b$, to be read off from the scaling behaviour (\ref{1.2}) of the
autocorrelator, is $b=2\beta/(\nu z)$ for a critical magnet and $b=-2\beta$ for interfaces. 
This is consistent with the scaling behaviour of the equal-time variance/interface width, respectively. 
On the other hand, for magnets relaxing towards an equilibrium state, a fluctuation-dissipation
theorem gives $a=b=2\beta/(\nu z)$, see e.g. \cite{Henk10}, 
whereas no such generic result is known for the fundamentally non-equilibrium 
interfaces.\footnote{In the {\sc ew} model, one has $a=b$ \cite{Roet06}, 
while in the $1D$ {\sc kpz} model $1+a=b+2/z$ \cite{Henk10} and the relationship for $d>1$ is unresolved 
\cite{Odor14}, see table~\ref{tab1}.} The analogy between magnetic systems and interfaces is further 
illustrated by the form of the kinetic equations of
the time-dependent order parameter, schematically written as 
$\partial_t\phi = -D\delta{\cal H}/\delta\phi+\eta$, where $\eta$ is a gaussian white noise. 
Clearly, the gaussian field/{\sc ew} model share mean-field characteristics, whereas the behaviour 
of the Ising and {\sc kpz} models is determined by the non-linearities in their equations of motion.
These non-linearities make exact solutions so difficult to find: for instance, in the $2D$ 
Ising model at equilibrium, only the cases (i) of a vanishing
magnetic field $h=0$ and arbitrary temperature $T$ \cite{Onsa44} and (ii) of the fixed temperature 
$T=T_c$ and an arbitrary magnetic field $h$ \cite{Zamo89,Warn94} have been solved (see \cite{Bort11} for a
recent overview of numerical and experimental tests).  

In magnetic phase transitions, in view of these technical difficulties, the so-called 
{\em spherical model} was originally proposed by Berlin and Kac
\cite{Berl52} in order to be able to explore generic properties of equilibrium phase transitions 
in  the context of a non-trivial exactly solvable model. 
A simple way to introduce it is to replace the discrete Ising spins $\sigma_i=\pm 1$, 
attached to each site $i$ of the lattice $\Lambda$, by continuous spin variables
$\sigma_i \mapsto s_i\in\mathbb{R}$ subject to the `spherical constraint' 
$\sum_{i\in\Lambda} s_i^2 \stackrel{!}{=} {\cal N}$, where $\cal N$ is the number of sites of the lattice. 
Especially in the form of the `mean spherical model' (where the spherical constraint is 
only required on average \cite{Lewi52}, which considerably
shortens the calculations, without modifying the critical behaviour), the 
spherical model has become an often-used test case in many
distinct situations. For the most common case of short-ranged interactions, 
a phase transition with a critical temperature $T_c>0$ exists in 
dimensions $d>2$. For dimensions $2<d<4$, the equilibrium exponents are different 
from the mean-field theory of the free gaussian field \cite{Berl52}. 
For the dynamics, and in the continuum limit, this amounts in the kinetic equation of motion to replace the non-linearity by a more simple
term, viz. $\phi^3 \mapsto  \left\langle \phi^2\right\rangle\phi =:  \mathfrak{z}(t)\phi$ 
and to fix the Lagrange multiplier $\mathfrak{z}(t)$ through the spherical constraint. 
This procedure can be applied to study the relaxational dynamics and to extract the exponents of 
both equilibrium and non-equilibrium critical dynamics, and was performed many times, see e.g. 
\cite{Ronc78,Coni94,Cugl94,Cugl95,Godr00b,Zipp00,Cann01,Fusc02,Pico02,Paes03,Sire04,Anni06,Cham06,Hase06,Baum06b,Baum07,Baum07b,Dutt08,Ebbi08,Fort12,Hase12,Godr13} 
and references therein. 
For reviews, see \cite{Joyc72,Bray00,Godr02,Cugl09,Henk10,Taeu14}. 

Here, we shall inquire whether a spherical model variant can be sensibly defined for 
models describing growing interfaces. In particular, we shall discuss the following questions:
\begin{enumerate}
\item which property of the interface heights could be construed to take only values $\pm 1$ 
in order to identify a sensible `spherical approximation'~? 
\item does one obtain a non-trivial model, distinct from the mean-field-like {\sc ew} class ? 
And does there exist\footnote{The existence of a finite $d^*$ would be analogous to critical magnets, 
whereas for the {\sc kpz} equation the question is
still unresolved, although numerical results suggest that $d^*$ might be infinite, see \cite{Alve14,Kim14,Rodr15}.} 
a well-defined upper critical dimension $d^*$~? 
\item in spite of the analogy with magnetic systems relaxing towards equilibrium, 
is the relaxation process an equilibrium or a non-equilibrium one~?
\end{enumerate}

This work is organised as follows: section~2 defines the Arcetri model and section~3 outlines the exact calculation of the interface width and 
of the two-time responses and correlators. The $1D$ model is shown to be equivalent to the $p=2$ spherical spin glass.  In section~4, the
long-time behaviour is explicitly found. We discuss in detail the non-equivalence with the {\sc ew}-model for all dimensions $d\ne 2$;
the relationship of the relaxation behaviour with the one of a bosonic particle-reaction model for dimensions $2<d<4$; and comment on the
global persistence probability. 
\textcolor{black}{We conclude in section~5.}  
Details of the calculations are treated in appendix~A. The Yeung-Rao-Desai inequality is revisited in appendix~B. 

\section{The Arcetri model}

We begin by stating the definition of the model\footnote{The name is inspired by the {\em lieu} 
where this model was conceived and this work was done.} 
to be analysed, first in $d=1$ dimensions. Consider a set of height variables $h_n(t)$ attached to the sites $n$ 
of a ring with $N$ sites. In what follows, we shall work with its (discrete, symmetrised) derivatives
\BEQ \label{2.1}
u_n(t) := \demi \left( h_{n+1}(t) - h_{n-1}(t) \right)
\EEQ
Furthermore, to each lattice site one attaches a gaussian random variable $\eta_n(t)$, with the moments
\BEQ\label{2.2}
\left\langle \eta_n(t) \right\rangle =0 \;\; , \;\; 
\left\langle \eta_n(t) \eta_m(t') \right\rangle = 2 \Gamma T \delta(t-t') \delta_{n,m}
\EEQ
The defining equations of motion for the $1D$ Arcetri model are
\BEA
\partial_t u_n(t) &=& \Gamma \left( u_{n+1}(t) + u_{n-1}(t) - 2 u_n(t) \right) + \mathfrak{z}(t) u_n(t) 
+ \demi \left( \eta_{n+1}(t) - \eta_{n-1}(t) \right)
\label{2.3} \\
\sum_{n=0}^{N-1} \left\langle u_n(t)^2 \right\rangle &=& N 
\label{2.4}
\EEA
where the Lagrange multiplier\footnote{We do not consider any fluctuations in $\mathfrak{z}(t)$, since they do not contribute to the spatially
local averages \cite{Anni06} on which we concentrate here.} $\mathfrak{z}(t)$ is determined from the `spherical constraint' (\ref{2.4}) 
and $\Gamma$ and $T$ are constants. 

\begin{figure}[tb]
\centerline{\psfig{figure=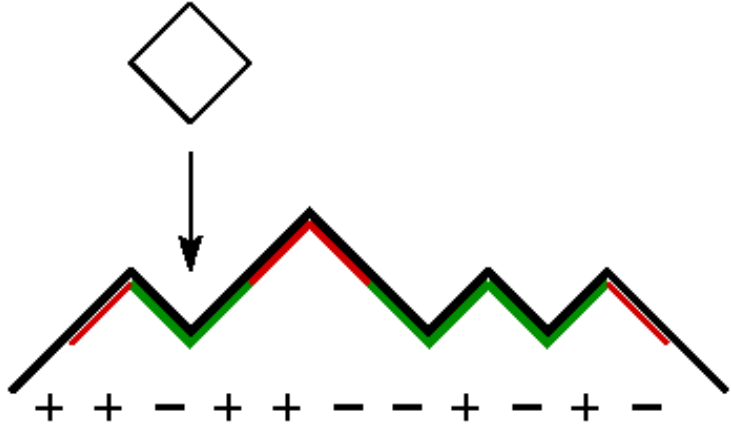,width=3.8in,clip=}}
\caption[fig1]{Schematic illustration of a $1D$ growing interface respecting the {\sc rsos} constraint that
the local slopes $u_{i+1/2} = h_{i+1}-h_{i}\stackrel{!}{=}\pm 1$ on nearest-neighbour sites. 
The interface grows through sequential absorption of square-shaped particles. 
Those `active' positions into which an adsorption is possible
are underlined in green and those `inactive' ones where this is prohibited by the 
{\sc rsos} constraint are underlined in red.
Active and inactive positions have to be updated after each adsorption process. \label{fig1}
}
\end{figure}
Our motivation to study this model comes from the well-known representation of lattice realisations of the 
{\sc kpz} universality class in terms of the totally asymmetric exclusion model ({\sc tasep}) 
\cite{Kard86,Daqu11,Taeu14}. In figure~\ref{fig1} an interface is sketched, and its growth
occurs through the adsorption of additional blocks, one at a given time, but such that between neighbouring sites 
the {\sc rsos}-condition $h_{i+1}(t)-h_i(t)=\pm 1$ is obeyed. Below the height configuration, 
the value of the interface $u_{i+1/2}=h_{i+1}-h_i$ on the dual lattice (the links)
is indicated. Let $u_{i+1/2}=-1 \mapsto \bullet$ represent a dual site occupied by a particle and 
$u_{i+1/2}=+1 \mapsto \circ$ an empty dual site. Then the adsorption process corresponds to an 
irreversible change $\bullet\circ \to \circ\bullet$ of a particle hopping to the right between
two neighbouring sites of the dual lattice, whereas the inverse reaction $\circ\bullet\to\bullet\circ$ 
is not admitted. In the continuum limit, this process is described by the {\sc kpz} equation, 
as given in table~\ref{tab2}.\footnote{\textcolor{black}{See \cite{Bert97} for a rigorous proof in $1D$.} 
Another often-used lattice representation is the Kim-Kosterlitz model \cite{Kim89}, 
where $h_{i+1}-h_i=\pm 1,0$, such that the continuum limit is again the {\sc kpz} equation.} 

In our search for a spherical model analogue for the {\sc kpz} equation, 
we shall therefore identify the {\em slopes} (\ref{2.1}) 
as the variables on which one could impose a spherical condition (\ref{2.4}). 
Then, it should be preferable to consider the equation of motion
of the slope, which in the continuum limit becomes $u(t,x) = \partial_x h(t,x)$ and obeys the noisy Burgers equation
\BEQ \label{2.5}
\partial_t u(t,x) = \nu \partial_x^2 u(t,x) + \mu u(t,x) \partial_x u(t,x) + \partial_x \eta(t,x)
\EEQ
Finally, we replace the non-linearity $u\textcolor{black}{\partial_x} u \mapsto \mathfrak{z}(t) u$ in (\ref{2.5}) 
by a spherical model construction, analogously
to the definition of the spherical model of a ferromagnet \cite{Berl52,Lewi52} and fix the
Lagrange multiplier $\mathfrak{z}(t)$ from the spherical constraint 
(\ref{2.4}).\footnote{In the kinetic spherical model of a ferromagnet, 
using a mean spherical constraint simplifies considerably the calculations, \cite{Godr00b}. Provided the infinite-system
limit $N\to\infty$ is taken before the large-time limit,
the consideration of a non-averaged spherical constraint leads to the same results for the long-time behaviour  \cite{Fusc02}.}
Eqs.~(\ref{2.3}),(\ref{2.4}) give the discretised version of this construction. 

\textcolor{black}{Clearly, eqs.~(\ref{2.4}),(\ref{2.5}) are identical to those of the conventional spherical model, 
see e.g. \cite{Godr00b,Pico02}, but with the standard
white noise $\eta(t,x)$ replaced by its derivative $\partial_x \eta(t,x)$. This new feature brings down the upper critical dimension
from $d^*=4$ to $d^*=2$ \cite{Bara95}. It is more surprising to find that the mere modification of the noise is enough to render our model identical to
the spherical spin glass \cite{Cugl95}, as we shall show in section~2.4. On the other hand, the present work with its specific way of
introducing the spherical constraint is merely meant as a first step in a more systematic exploration of `spherical variants' of equations 
such as (\ref{2.5}). We shall come back in section~5
to other possibilities which might keep a more clear memory of the non-linear terms in (\ref{2.5}). Higher-order non-linearities will either
turn out to be irrelevant or else may create some `multicritical' variant of scale-invariant growth \cite{Bara95,Halp95,Krug97}.} 

In order to see how to generalise this model to any dimension $d\ne 1$, we briefly consider the $2D$ case, for simplicity of notation. 
The height on each site is $h_{n,m}(t)$. Then we have slopes
in each of the two lattice directions, namely
\BEQ\label{2.6}
u_{n,m}(t) := \demi \left( h_{n+1,m}(t) - h_{n-1,m}(t) \right) \;\; , \;\; 
v_{n,m}(t) := \demi \left( h_{n,m+1}(t) - h_{n,m-1}(t) \right)
\EEQ
which obey the equations of motion (for brevity, the dependence on $t$ is suppressed)
\BEA
\partial_t u_{n,m} &=& \Gamma \left( u_{n+1,m} + u_{n-1,m} + u_{n,m+1} + u_{n,m-1} - 4 u_{n,m}\right) 
+ \mathfrak{z}(t) u_{n,m} +\demi \left( \eta_{n+1,m} - \eta_{n-1,m} \right) 
\nonumber \\
\partial_t v_{n,m} &=& \Gamma \left( v_{n+1,m} + v_{n-1,m} + v_{n,m+1} + v_{n,m-1} - 4 v_{n,m}\right) 
+ \mathfrak{z}(t) v_{n,m} +\demi \left( \eta_{n,m+1} - \eta_{n,m-1} \right) 
\nonumber \\
\label{2.7}
\EEA
Considering the spherical constraint, we must count how many links exist 
on a torus $\mathbb{S}^1\otimes \mathbb{S}^1$ with $N\times N=N^2$ sites. 
Since on each line of the lattice, one has $N$ links, there are $N$ 
such lines and there two dimensions along with the links must be counted, 
there is a total of $N\times N \times 2 = 2 N^2$ links. Hence, the spherical constraint must be
\BEQ \label{2.8}
\sum_{n,m=0}^{N-1} \left\langle \left( u_{n,m}^2 + v_{n,m}^2 \right) \right\rangle = 2 N^2
\EEQ
The extension to generic $d$ is now obvious and will be written down explicitly in the next section.\footnote{The formulation chosen here is
explicitly invariant under spatial rotations. Any attempt of a generalisation towards more than one spherical constraint will likely lead to
a breaking of that symmetry.}  

Clearly, in analogy with the {\sc ew} and {\sc kpz} equations, we have already performed implicitly a Galilei transformation and shall
always work in a reference frame which moves upwards with the average rate of particle deposition. The parameter $T$ in the noise correlator
serves to distinguish the kinetic noise from the effective diffusion constant $\nu$. 

\section{Solution}

\subsection{Equations of motion}

The solution of the Arcetri model equations of motion, i.e. eqs.~(\ref{2.7}),(\ref{2.8}) for $d=2$, is carried out in Fourier space. In $d$ spatial
dimensions, there are $d$ slopes, denoted $u_{a,\vec{n}}(t)$, with $a=1,\ldots,d$ and $\vec{n}=(n_1,\ldots,n_d)$. On a torus 
$\mathbb{T}^d = \mathbb{S}^1\otimes \cdots \otimes \mathbb{S}^1 = \left(\mathbb{S}^1\right)^{\otimes d}$, 
represented by a hyper-cubic lattice with $N^d$ sites, one has the Fourier transforms
\BEA
\wht{u}_a(t,\vec{p}) &=& \sum_{n_1=0}^{N-1}\cdots\sum_{n_d=0}^{N-1} 
\exp\left( -\frac{2\pi\II}{N} \vec{p}\cdot\vec{n}\right) u_{a,\vec{n}}(t) 
\nonumber \\
u_{a,\vec{n}}(t) &=& \frac{1}{N^d} \sum_{p_1=0}^{N-1}\cdots\sum_{p_d=0}^{N-1} 
\exp\left(\frac{2\pi\II}{N} \vec{p}\cdot\vec{n}\right) \wht{u}_a(t,\vec{p})
\label{3.1}
\EEA
Our equations of motion take the form
\BEQ \label{3.2}
\partial_t \wht{u}_a(t,\vec{p}) = -2\Gamma\omega(\vec{p}) \wht{u}_a(t,\vec{p})+ \mathfrak{z}(t)\wht{u}_a(t,\vec{p})
+\II\sin\left(\frac{2\pi}{N} p_a\right) \wht{\eta}(t,\vec{p})
\EEQ
with the dispersion relation $\omega(\vec{p}) = \sum_{j=1}^d \left[1-\cos\left(\frac{2\pi}{N}p_j\right)\right]$, 
and the gaussian noise correlators
\BEQ
\left\langle \wht{\eta}(t,\vec{p})\right\rangle =0 \;\; , \;\;
\left\langle \wht{\eta}(t,\vec{p})\wht{\eta}(t',\vec{q})\right\rangle = 2\Gamma T N^d \delta(t-t') \delta_{\vec{p}+\vec{q},\vec{0}}
\EEQ
Since the definition (\ref{2.6}) implies integrability conditions,\footnote{\textcolor{black}{From (\ref{2.6}), one has in $2D$ that
$\demi (u_{n,m+1}(t)-u_{n,m-1}(t))=\demi(v_{n+1,m}(t)-v_{n-1,m}(t))$, which in the continuum limit amounts to 
$\partial u/\partial y=\partial v/\partial x$. Standard calculus \cite[p. 104]{Cour74} then shows that there exists a function 
$h$ such that $u=\partial h/\partial x$ and $v=\partial h/\partial y$, on a simply connected domain. Eq.~(\ref{3.4}) expresses this in
discrete Fourier space.}} 
the interface height is given by
\BEQ \label{3.4}
\wht{u}_a(t,\vec{p}) = \II \sin\left(  \frac{2\pi}{N} p_a\right)  \wht{h}(t,\vec{p}) \mbox{\rm ~~;~~ if $\vec{p}\ne\vec{0}$,~ $a=1,\ldots,d$}
\EEQ
We restrict attention to interfaces which \textcolor{black}{on average} are initially flat and uncorrelated. 
\textcolor{black}{Indeed, the initial modes} $\wht{h}(0,\vec{p})$ are assumed gaussian random variables with the moments
\BEQ \label{3.5}
\left\langle \wht{h}(0,\vec{p})\right\rangle = N^d H_0\: \delta_{\vec{p},\vec{0}} \;\; , \;\;
\left\langle \wht{h}(0,\vec{p})\wht{h}(0,\vec{q})\right\rangle = N^{d} H_1\: \delta_{\vec{p}+\vec{q},\vec{0}}
\EEQ
where $H_{0,1}$ are initial parameters. 
\textcolor{black}{The first condition (\ref{3.5}) means that the average initial slopes vanish. From the second, if} 
$H_1=H_0^2$, the initial interface width vanishes. \textcolor{black}{We shall restrict to this case below, see eq.~(\ref{3.22B}).} 

The equations of motion (\ref{3.2}) have the following solution, with $a=1,\ldots,d$
\BEA
\wht{u}_a(t,\vec{p}) &=& \wht{u}_{a}(0,\vec{p}) \exp\left[ -2\Gamma \omega(\vec{p}) t +\int_0^t \!\D t'\: \mathfrak{z}(t')\right] 
\nonumber \\
& & +\int_0^t \!\D\tau\: \II\sin\left(\frac{2\pi}{N}p_a\right) \wht{\eta}(\tau,\vec{p}) 
\exp\left[ -2\Gamma \omega(\vec{p}) (t-\tau) +\int_{\tau}^t \!\D t'\: \mathfrak{z}(t')\right]
\label{3.6}
\EEA
In order to work out the explicit form of the spherical constraint, 
we follow the lines of the magnetic spherical model and define \cite{Godr00b}
\BEQ
g(t) := \exp\left( -2 \int_0^t \!\D t'\: \mathfrak{z}(t) \right)
\EEQ
Then the spherical constraint becomes, using (\ref{3.1}) and the initial condition (\ref{3.4}),(\ref{3.5}), 
\BEA
d N^d &\stackrel{!}{=}& \sum_{a=1}^d \sum_{n_1=0}^{N-1} \cdots \sum_{n_d=0}^{N-1} 
\left\langle u_{a,\vec{n}}(t)^2 \right\rangle 
\nonumber \\
&=& \sum_{p_1=0}^{N-1}\cdots \sum_{p_d=0}^{N-1} \left[ H_1 \lambda(\vec{p}) e^{-4\Gamma \omega(\vec{p}) t} \frac{1}{g(t)} 
 + \int_0^t \!\D\tau\: 2\Gamma T\: \lambda(\vec{p}) e^{-4\Gamma \omega(\vec{p}) (t-\tau)} \frac{g(\tau)}{g(t)} \right]
\EEA
with $\lambda(\vec{p}) := \sum_{a=1}^d \sin^2\left(\frac{2\pi}{N} p_a\right)$. If we now define
\BEQ \label{3.9}
f(t) := \frac{1}{N^d} \sum_{p_1=0}^{N-1}\cdots\sum_{p_d=0}^{N-1} \lambda(\vec{p})\, e^{-4\Gamma t \omega(\vec{p})}
\EEQ
the spherical constraint can be written as a Volterra integral equation
\BEQ \label{3.10}
H_1\, f(t) + 2\Gamma T \int_0^t \!\D\tau\: g(\tau) f(t-\tau) = d\, g(t)
\EEQ
Its solution will be found in complete analogy with well-established techniques \cite{Ronc78,Cugl95,Godr00b}. 

At this point, the infinite-size limit  $N\to\infty$ can be taken. The solution (\ref{3.6}) becomes
\BEQ \label{3.11}
\wht{u}_a(t,\vec{p}) = \wht{u}_a(0,\vec{p}) e^{-2\Gamma t \omega(\vec{p})} \frac{1}{\sqrt{g(t)}} 
+\int_0^t \!\D\tau\: \II \sin p_a\: \wht{\eta}(\tau,\vec{p}) e^{-2\Gamma (t-\tau)\omega(\vec{p})} \sqrt{\frac{g(\tau)}{g(t)}}
\EEQ
where the dispersion relation now reads $\omega(\vec{p}) = \sum_{a=1}^d \left[ 1 - \cos p_a\right]$ 
and $\vec{p}\in {\cal B}:=[-\pi,\pi]^d$ is in the Brillouin zone. Letting now
$\lambda(\vec{p}):= \sum_{a=1}^d \sin^2 p_a$, we have from (3.9), 
\BEA
f(t) &=& \frac{1}{(2\pi)^d} \int_{\cal B} \!\D\vec{p}\: \lambda(\vec{p})\, e^{-4\Gamma t \omega(\vec{p})}
\nonumber \\
&=& \frac{d}{2\pi} \int_{-\pi}^{\pi} \!\D p\: \sin^2 \!p~ \; e^{4\Gamma t \cos p}\: 
\left( \frac{1}{2\pi}\int_{-\pi}^{\pi} \!\D q\: e^{4\Gamma t\cos q} \right)^{d-1}
e^{-4d\Gamma t}
\nonumber \\
&=& d\,\frac{e^{-4\Gamma t} I_1(4\Gamma t)}{4\Gamma t} \left( e^{-4\Gamma t} I_0(4\Gamma t)\right)^{d-1}
\label{3.12}
\EEA
where the $I_n$ denote modified Bessel functions \cite{Abra65}. The function $g(t)$ is found from the Volterra integral equation (\ref{3.10}). 
Furthermore, since (\ref{3.4}) is now written as $\wht{u}_a(t,\vec{p}) = \II \sin p_a\, \wht{h}(t,\vec{p})$, if $\vec{p}\ne \vec{0}$, 
the interface height becomes\footnote{The {\sc ew} model is obtained by setting $g(t)=1 \leftrightarrow \mathfrak{z}(t)=0$ \cite{Roet06}.}
\BEQ \label{3.13}
\wht{h}(t,\vec{p}) = \wht{h}(0,\vec{p})\, e^{-2\Gamma t \omega(\vec{p})} \frac{1}{\sqrt{g(t)}}
+\int_0^t \!\D\tau\:  \wht{\eta}(\tau,\vec{p})\, e^{-2\Gamma (t-\tau)\omega(\vec{p})} \sqrt{\frac{g(\tau)}{g(t)}\,}
\EEQ
and the initial conditions (\ref{3.5}) now read
\BEQ \label{3.14}
\left\langle \wht{h}(0,\vec{p})\right\rangle = H_0\: \delta(\vec{p}) \;\; , \;\;
\left\langle \wht{h}(0,\vec{p})\wht{h}(0,\vec{q})\right\rangle = H_1\: \delta(\vec{p}+\vec{q})
\EEQ
The calculation of the long-time behaviour of physical observables will be based on eqs.~(\ref{3.11}),(\ref{3.13}), respectively. To do this
in practice, $g(t)$ must be found from the constraint (\ref{3.10}), with the explicit form (\ref{3.12}) of $f(t)$ taken into account. 

\subsection{Linear responses}

Since the evolution of the model starts from a non-stationary initial state and the dynamics contains 
an external force which does not derive from a Hamiltonian, it leads to non-equilibrium relaxation, 
where responses and correlators must be analysed separately. We begin with the response functions. 

As the Arcetri model describes the growth of an interface, the conjugate variable 
to the interface height $\wht{h}$ is an additional particle-deposition
rate $\wht{\jmath}=\wht{\jmath}(s,\vec{q})$ at time $s$ and at momentum $\vec{q}$. 
Such an extra term can simply be added to the equations of motions by formally
replacing $\eta \mapsto \eta + j$. We then define the linear response of the interface height
\BEQ
\wht{R}(t,s;\vec{p},\vec{q}) := \left. \frac{\delta \langle \wht{h}(t,\vec{p})\rangle}{\delta \wht{\jmath}(s,-\vec{q})}\right|_{j=0} 
=\Theta(t-s) \delta(\vec{p}-\vec{q}) \sqrt{\frac{g(s)}{g(t)}\,}\: e^{-2\Gamma (t-s)\omega(\vec{p})}
\EEQ
and used (\ref{3.13}). For a linear response as considered here, $g(t)$ is still given by (\ref{3.10}). 
The Heaviside function $\Theta(t-s)$ expresses the causality condition $t>s$. In direct space, this gives
\BEA
\lefteqn{ R(t,s;\vec{r}-\vec{r}') = \left. \frac{\delta \langle {h}(t,\vec{r})\rangle}{\delta j(s,\vec{r}')}\right|_{j=0} }
\nonumber \\
&=& 
\frac{\Theta(t-s)}{(2\pi)^d} \sqrt{\frac{g(s)}{g(t)}\,}\: \int_{\cal B} \!\D\vec{p}\; 
e^{\II \vec{p}\cdot(\vec{r}-\vec{r}') -2\Gamma (t-s)\omega(\vec{p})} 
\:=\: \Theta(t-s) \sqrt{\frac{g(s)}{g(t)}\,}\: F_{\vec{r}}(t-s)
\label{3.16}
\EEA
where here and in what follows we use the abbreviation (with the modified Bessel function $I_n$ taken from \cite[eq. (9.6.19)]{Abra65})
\BEQ \label{3.22}
F_{\vec{r}}(\tau) := \frac{1}{(2\pi)^d} \int_{\cal B}\!\D\vec{p}\: \cos(\vec{p}\cdot\vec{r})\, e^{-2\Gamma \tau\, \omega(\vec{p})}
= \prod_{a=1}^d e^{-2\Gamma\tau} I_{r_a}(2\Gamma \tau)
\EEQ

In order to derive the response of the slopes, consider first a source term in an action, expressed in momentum space 
\BD
d\int \!\D\vec{p}\: \wht{h}(t,\vec{p}) \wht{\jmath}(t,-\vec{p}) 
= \int\!\D\vec{p}\: \sum_{a=1}^d \wht{h}(t,\vec{p}) \frac{\II\sin p_a}{\II\sin p_a}\, \wht{\jmath}(t,-\vec{p})
= \sum_{a=1}^d \int\!\D\vec{p}\:  \wht{u}_a(t,\vec{p}) \wht{J}_a(t,-\vec{p})
\ED
using (\ref{3.4}) and where we introduced the integrated source $\wht{J}_a(t,\vec{p}) := (-\II \sin p_a)^{-1}\, \wht{\jmath}(t,\vec{p})$. 
Consider the following linear response 
\BEQ
\wht{Q}(t,s;\vec{p},\vec{q}) := \sum_{a=1}^d \left. \frac{\delta \langle \wht{u}_a(t,\vec{p})\rangle}{\delta \wht{J}_a(s,-\vec{q})}\right|_{j=0} 
=\delta(\vec{p}-\vec{q}) \Theta(t-s) \sqrt{\frac{g(s)}{g(t)}\,}\: \sum_{a=1}^d \sin^2\! p_a\: e^{-2\Gamma (t-s)\omega(\vec{p})}
\EEQ
which in direct space leads to 
\BEA
Q(t,s;\vec{r}-\vec{r}') &=& 
\frac{1}{(2\pi)^{2d}}\int_{{\cal B}^2} \!\D\vec{p}\,\D\vec{q}\: e^{\II \vec{p}\cdot(\vec{r}-\vec{r}')} \wht{Q}(t,s;\vec{p},\vec{q})
\nonumber \\
&=& \frac{\Theta(t-s)}{(2\pi)^d} \sqrt{\frac{g(s)}{g(t)}\,}\, \int_{\cal B} \!\D\vec{p}\: \lambda(\vec{p})\: 
e^{\II\vec{p}\cdot(\vec{r}-\vec{r}')-2\Gamma (t-s)\omega(\vec{p})} 
\label{3.18}
\EEA
In particular, the autoresponse of the slopes takes the simple form $Q(t,s) = Q(t,s;\vec{0}) = \Theta(t-s) f((t-s)/2) \sqrt{g(s)/g(t)}$. 

\subsection{Correlation functions}

In order to write down the fluctuations and correlations of the height, we average the solution (\ref{3.13}) for $\vec{p}\ne\vec{0}$. Then
\BEQ
\left\langle \wht{h}(t,\vec{p})\right\rangle = \left\langle \wht{h}(0,\vec{p})\right\rangle \frac{\exp(-2\Gamma t\omega(\vec{p}))}{\sqrt{g(t)}\,}
\EEQ
Going back to discrete momenta formulation for clarity, we have the deviation in the height
\BEA
h(t,\vec{r}) - \langle h(t,\vec{r})\rangle &=& 
\frac{1}{N^d} \sum_{\vec{p}} \left[ \left( \wht{h}(0,\vec{p}) - \langle\wht{h}(0,\vec{p})\rangle \right)
\exp\left(\frac{2\pi\II}{N}\vec{p}\cdot\vec{r} -2\Gamma t\omega(\vec{p})\right) \frac{1}{\sqrt{g(t)}\,} \right.
\nonumber \\
& & \left. +\int_0^t \!\D\tau\: \wht{\eta}(\tau,\vec{p}) 
\exp\left(\frac{2\pi\II}{N}\vec{p}\cdot\vec{r} -2\Gamma (t-\tau)\omega(\vec{p})\right) \sqrt{\frac{g(\tau)}{g(t)}\,}\; \right]
\EEA
such that the two-time height-height correlator becomes
\BEA
\lefteqn{C(t,s;\vec{r}-\vec{r}')  = C(s,t;\vec{r}'-\vec{r}) = 
\left\langle \left( h(t,\vec{r}) - \langle h(t,\vec{r}) \rangle\right)\left( h(s,\vec{r}') - \langle h(s,\vec{r}') \rangle \right) \right\rangle}
\nonumber \\
&=& \frac{1}{N^d}\frac{1}{\sqrt{g(t)g(s)\,}} \sum_{\vec{p}\ne\vec{0}} \left[ 
\left\langle \wht{h}(0,\vec{p})\wht{h}(0,-\vec{p})\right\rangle - \left\langle \wht{h}(0,\vec{p})\right\rangle^2\right]\: 
e^{2\pi\II N^{-1} \vec{p}\cdot(\vec{r}-\vec{r}')}\, e^{-2\Gamma (t+s)\omega(\vec{p})}
\nonumber \\
& & + \frac{2\Gamma T}{N^d} \sum_{\vec{p}} \int_0^{\min(t,s)}\!\D\tau\: \frac{g(\tau)}{\sqrt{g(t)g(s)\,}\,}\: 
e^{2\pi\II N^{-1} \vec{p}\cdot(\vec{r}-\vec{r}')}\, e^{-2\Gamma (t+s-2\tau)\omega(\vec{p})}
\label{3.22B}
\EEA
For a vanishing initial width, see eq.~(\ref{3.5}) with $H_1=H_0^2$, 
the term in the square brackets in the 2$^{\rm nd}$ line vanishes for 
$\vec{p}=\vec{0}$. We can now take again the $N\to\infty$ limit. Using $F_{\vec{r}}(t)$ as defined in eq.~(\ref{3.22}), 
the two-time correlator reads (simplified for $t\geq s$)
\BEQ \label{3.23}
C(t,s;\vec{r}) = \frac{H_1}{\sqrt{g(t)g(s)\,}\,}\, F_{\vec{r}}(t+s) 
+ \frac{2\Gamma T}{\sqrt{g(t)g(s)\,}\,}\, \int_{0}^{s}\!\D\tau\: g(\tau) F_{\vec{r}}(t+s-2\tau)
\EEQ
and the autocorrelator becomes $C(t,s)=C(t,s;\vec{0})$. In particular, the interface width reads
\BEQ \label{3.24}
w^2(t) = C(t,t) = \frac{H_1 F_{\vec{0}}(2t)}{g(t)} + 2\Gamma T \int_0^t \!\D\tau\: \frac{g(\tau)}{g(t)} F_{\vec{0}}(2t-2\tau)
\EEQ

Similarly, the slope-slope correlator can be found, again with $t\geq s$ assumed 
\BEA
\left\langle u_a(t,\vec{r}) u_a(s,\vec{r}')\right\rangle &=& 
\frac{H_1}{\sqrt{g(t)g(s)\,}\,} \frac{1}{(2\pi)^d} \int_{\cal B} \!\D\vec{p}\: \sin^2 p_a\:
e^{\II\vec{p}\cdot(\vec{r}-\vec{r}') -2\Gamma (t+s)\omega(\vec{p})} 
\\
& & + \frac{2\Gamma T}{(2\pi)^d} \int_{\cal B}\!\D\vec{p}\int_0^{s}\!\D\tau\: \frac{g(\tau)}{\sqrt{g(t)g(s)}\,}\, \sin^2 p_a\:
e^{\II\vec{p}\cdot(\vec{r}-\vec{r}') -2\Gamma (t+s-2\tau)\omega(\vec{p})} 
\nonumber
\EEA
whereas $\langle u_a u_b\rangle=0$ if $a\ne b$. 
We shall be interested in particular in the autocorrelator of the slopes, with $f(t)$ given by eq.~(\ref{3.12}) 
\BEQ \label{3.26}
A(t,s) := \sum_{a=1}^d \left\langle u_a(t,\vec{r}) u_a(s,\vec{r})\right\rangle 
= H_1 \frac{f((t+s)/2)}{\sqrt{g(t)g(s)}\,} 
+ 2\Gamma T \int_0^{s}\!\D\tau\: \frac{g(\tau)}{\sqrt{g(t)g(s)}\,}\, f\left(\frac{t+s}{2}-\tau\right)
\EEQ
Here, and in what follows, we shall always choose units such that $\Gamma=1$.

\subsection{Equivalence with the spherical spin glass}

As we now show, {\em the $1D$ Arcetri model is equivalent to the kinetic 
$p=2$ spherical spin glass}, analysed by Cugliandolo and Dean \cite{Cugl95}.\footnote{A careful analysis  
\cite{Cham06,Cham11} shows that the ageing properties of truly glassy systems
on one hand, and simple spin systems and the $p=2$ spherical spin glass on the other hand, are quite distinct.}

The spin hamiltonian is ${\cal H}=-\demi\sum_{i\ne j} J_{i,j} s_i s_j$, where 
$s_i\in\mathbb{R}$ are spherical spins which obey the spherical
constraint $\sum_{i=1}^N s_i^2 = N$. The elements of the symmetric matrix $J$ 
are independent gaussian random variables with zero mean and variance proportional
to $1/N$. Then the thermodynamic limit is well-defined and the probability 
distribution of the eigenvalues $\mu$ of $J$ is given by the Wigner semi-circle law
\BEQ
\rho(\mu) = \frac{\sqrt{4-\mu^2\,}}{2\pi}
\EEQ
The corresponding eigenvectors of a spin configuration $s(t) = \{ s_i(t)\}$ 
are denoted by $s_{\mu}(t) = \mu\cdot s(t)$. Consider uniform initial conditions
$s_{\mu}(0)=1$. The time-evolution is given by the Langevin equation \cite[eq. (2.2)]{Cugl95}
\BEQ \label{3.28}
\partial_t s_{\mu}(t) = \left( \mu +\mathfrak{z}(t)\right) s_{\mu}(t) + h_{\mu}(t) + \xi_{\mu}(t)
\EEQ
where $h_{\mu}(t)$ is an external magnetic field and $\xi_{\mu}(t)$ is the centered thermal noise, with variance 
$\langle \xi_{\mu}(t)\xi_{\nu}(t')\rangle = 2 T_{\rm SG} \delta_{\mu\nu}\delta(t-t')$. 
The solution of the Langevin equation is now immediate. 
Defining $\gamma(t) := \exp(-2\int_0^t \!\D\tau\, \mathfrak{z}(\tau))$, 
the spherical constraint can be rewritten as a Volterra integral equation \cite[eq. (2.7)]{Cugl95}
\BEQ \label{3.29}
\gamma(t) = \left\langle\!\left\langle s_{\mu}(0) \exp(2\mu t)\right\rangle\!\right\rangle 
+ 2 T_{\rm SG} \int_0^t \!\D\tau\: \gamma(\tau) \left\langle\!\left\langle \exp 2\mu(t-\tau)\right\rangle\!\right\rangle
\EEQ
where the following average over the Wigner distribution $\rho(\mu)$ is carried out \cite{Cugl95}
\BEQ \label{3.30}
\left\langle\!\left\langle \exp(2\mu t)\right\rangle\!\right\rangle := \int_{-2}^2 \!\D\mu\: \rho(\mu) e^{2\mu t} 
= \frac{2}{\pi} \int_{-1}^1 \!\D\mu\: \sqrt{1-\mu^2\,}\: e^{4\mu t} = \frac{I_1(4t)}{2t}
\EEQ
where \cite[eq. (9.6.18)]{Abra65} was used. 

In order to see the relationship of these results with the $1D$ Arcetri model, 
we first observe that the eigenvalue spectrum in the spherical spin glass
is in the interval $\mu\in[-2,2]$, whereas the dispersion relation of the Arcetri model $2\omega(p)\in[0,4]$. 
This can be matched through the mapping $\mu \mapsto -2+\mu$ and $\mathfrak{z}(t)\mapsto \mathfrak{z}(t)+2$ 
in order to keep the equation of motion (3.28) unchanged. Identifying the relationship 
\BEQ
g(t) = e^{-4t} \gamma(t)
\EEQ 
it follows that {\em the Volterra equation (\ref{3.29}) with the explicit average (\ref{3.30}) 
becomes exactly the spherical constraint (\ref{3.10})}, where
$f(t)$ is given in (\ref{3.12}) and the identifications 
\BEQ \label{TH1}
T=2T_{\rm SG} \mbox{\rm ~~and~~} H_1=2
\EEQ 
Next, the magnetic autoresponse of the spin glass (with $t>s$ assumed) \cite[eqs. (2.16,3.17)]{Cugl95}
\BEQ \label{3.33}
R_{\rm SG}(t,s) := \sum_{i=1}^{N} \left. \frac{\delta \langle s_i(t)\rangle}{\delta h_i(s)}\right|_{h=0} = \sqrt{\frac{\gamma(s)}{\gamma(t)}\,}\,  
\left\langle\!\left\langle \exp \mu (t-s)\right\rangle\!\right\rangle = 2\sqrt{\frac{g(s)}{g(t)}\,}\, f\left(\frac{t-s}{2}\right) = 2Q(t,s)  
\EEQ
is identical to the slope autoresponse (\ref{3.18}), up to a factor 2. 
Finally, and recalling (\ref{3.30}), the disorder-averaged spin glass autocorrelator \cite[eqs. (2.12,3.6)]{Cugl95}
\BEA \label{3.34}
\lefteqn{C_{\rm SG}(t,s) := \frac{1}{N}\left[ \sum_{i=1}^N \left\langle s_i(t) s_i(s)\right\rangle\right]_J 
\:=\: \int_{-2}^2 \!\D\mu\: \rho(\mu) \left\langle s_{\mu}(t) s_{\mu}(s) \right\rangle} 
\\
&=& 
\frac{1}{\sqrt{\gamma(t)\gamma(s)}} \left( \left\langle\!\left\langle \exp\mu (t+s)\right\rangle\!\right\rangle 
+2T_{\rm SG} \int_0^{\min(t,s)} \!\!\D\tau\: \gamma(\tau) \left\langle\!\left\langle  
\exp \mu (t+s-2\tau)\right\rangle\!\right\rangle \right) =A(t,s)
\nonumber
\EEA
reduces to the slope-slope autocorrelator (\ref{3.26}), with the same identifications (\ref{TH1}) as above. 

Since it is well-known that the spherical spin glass is in the same universality 
class as the $3D$ kinetic spherical model \cite{Cugl95,Godr00b}, 
it follows that the $1D$ Arcetri model is in the same universality class as well. 
The physical correspondence is between the slopes in the 
Arcetri model and the magnetic spherical spins, see (\ref{3.33}),(\ref{3.34}). 
This can be generalised: the long-time asymptotics of $g(t)$ and $f(t)$ imply that 
{\em the exponents which
give the long-time behaviour of $A(t,s)$ and $Q(t,s)$ of the $d$-dimensional Arcetri model are the same as those of the
magnetic autocorrelators and autoresponses in the kinetic spherical model \cite{Godr00b} in $d+2$ dimensions}. 

\section{Long-time behaviour}

Given the form of the spherical constraint (\ref{3.10}) as a Volterra integral equation, 
its solution proceeds via standard Laplace transforms \cite{Ronc78,Cugl95,Godr00b}. 
Let $\lap{g}(p)={\cal L}(g(t))(p)=\int_0^{\infty}\!\D t\: g(t) e^{-pt}$ denote the Laplace transform. 
Then (\ref{3.10}) can be inverted
\BEQ \label{4.1}
\lap{g}(p) = \frac{H_1 \lap{f}(p)}{d-2T \lap{f}(p)}
\EEQ 
Standard Tauberian theorems \cite[ch. XIII.5]{Fell71} relate the long-time behaviour of $g(t)$ to the $p\to 0$ behaviour of $\lap{g}(p)$. 
Hence, expanding $\lap{f}(p)$ for $p\ll 1$, one can find first $\lap{g}(p)$ and then $g(t)$, for $t$ large enough.  
The explicit calculations are detailed in appendix~A. 

\subsection{Fast relaxation for $T>T_c$}

Consider the situation when the denominator in (\ref{4.1}) vanishes for some value $p_0>0$: $d-2T \lap{f}(p_0)=0$. 
Since $\lap{f}(p)$ decreases monotonously with $p$, this zero exists and is simple for $T$ large enough. 
Hence $\lap{g}(p)$ has a simple pole at $p=p_0$. Transforming back, one has
\BEQ
g(t) \stackrel{t\to\infty}{\simeq} - \frac{H_1}{2T}\frac{\lap{f}(p_0)}{{\lap{f}\,}'(p_0)}\, e^{ p_0 t} 
= - \frac{H_1d}{4T^2}\frac{1}{{\lap{f}\,}'(p_0)}\, e^{ p_0 t}
\EEQ
Here $1/p_0$ is the finite relaxation time and governs the time-translation-invariant approach towards the stationary state
($w_{0,1}$ are constants)
\BEA
R(s+\tau,s;\vec{r}) &\sim& \tau^{-d/2} \exp\left( - \demi p_0 \tau - \frac{\vec{r}^2}{4\tau}\right) \nonumber \\
w^2(t) &\sim& w_0 T + w_1 e^{- p_0 t} \\
C(s+\tau,s) &\sim& \exp\left( -\demi p_0 \tau\right) \nonumber
\EEA
This rapid relaxation does not show dynamical scaling and is of no particular interest to us. 

\subsection{Ageing at the critical point}

The smallest temperature $T=T_c$, for which $\lap{g}(p)$ has a singularity, occurs when $d/(2T_c) = \lap{f}(0)$. 
Hence the {\em critical temperature} $T_c=T_c(d)$ is given by
\BEQ \label{4.4}
\frac{1}{T_c(d)} = \frac{2}{d} \lap{f}(0) = \demi \int_0^{\infty} \!\D t\: e^{-dt} t^{-1} I_1(t) I_0(t)^{d-1}
\EEQ
Clearly, $T_c(d)>0$ for all $d>0$. 
Using \cite[eq. (11.4.13)]{Abra65} and \cite[eq. (2.15.20.6)]{Prudnikov2}, 
we have the explicit values\footnote{For $d=1$ in agreement with the
spin glass critical point $T_{{\rm SG},c}=1$ \cite{Cugl95}.}
\BEQ \label{4.5}
T_c(1) = 2 \;\; , \;\; T_c(2) = \frac{2\pi}{\pi-2} \simeq 5.5038\ldots \;\; , \;\; T_c(3) \simeq 9.53099\ldots 
\EEQ
In addition, we have $T_c(d)\simeq d$ for $d\ll 1$ and $T_c(d)\simeq 4d$ for $d\gg 1$.

We now write down the linear response (\ref{3.16}), the interface width (\ref{3.24}) and the autocorrelator (\ref{3.23}), 
in the long-time scaling limit, where both $t,s\to\infty$ such that $y =t/s$ is kept fixed. 
We begin with the two-time time-space response and expand $g(t)$ and $F_{\vec{r}}(t)$
for large times. This gives
\BEA
R(t,s;\vec{r}) &=& \sqrt{\frac{g(s)}{g(t)}\,}\, F_{\vec{r}}(t-s) \simeq R(t,s) \exp\left[ -\frac{1}{4}\frac{\vec{r}^2}{t-s}\right]
\nonumber \\
R(t,s) &=& s^{-d/2} f_R(t/s)
\label{4.6}
\EEA
with the explicit scaling function 
\BEA
f_R(y) &=& (4\pi)^{-d/2} \left( y-1\right)^{-d/2} y^{(2-d)/4} \mbox{\rm ~;~~ if $0<d<2$} \nonumber \\
f_R(y) &=& (4\pi)^{-d/2} \left( y-1\right)^{-d/2} \hspace{1.3truecm}\mbox{\rm ~;~~ if $2<d$}
\label{4.7}
\EEA
{}From this, the values of the non-equilibrium exponents $a,\lambda_R$ and also the dynamical exponent 
$z=2$ can be read off and are listed in table~\ref{tab1}. 

Next, consider the two-time correlator. In the scaling limit, with $t=ys > s$, and for $0<d<2$, we find
\BEA
C(t,s;\vec{r}) &\simeq& 2 T_c \left( t s\right)^{(2-d)/4} \int_0^s \!\D\tau\: \tau^{d/2-1} F_{\vec{r}}(t+s-2\tau)
\\
&\stackrel{s\to\infty}{\simeq}& s^{1-d/2}\, 2 T_c\, y^{(2-d)/4} \int_0^1 \!\D u\: u^{d/2-1} \left( 4\pi \left[ (y+1)-2u\right]\right)^{-d/2}
\exp\left[ - \frac{\vec{r}^2}{s} \frac{1}{y+1-2u}\right]
\nonumber
\EEA
which is of the expected scaling form $C(t,s;\vec{r}) = s^{-b} F_C(t/s,\vec{r}^2/s)$. For $d>2$, an analogous calculation gives
\BEQ 
C(t,s;\vec{r}) = s^{1-d/2}\, 2 T_c \int_0^1 \!\D u\: u^{d/2-1} \left( 4\pi \left[ (y+1)-2u\right]\right)^{-d/2}
\exp\left[ - \frac{\vec{r}^2}{s} \frac{1}{y+1-2u}\right]
\EEQ
Especially, the autocorrelator $C(t,s) = C(t,s;\vec{0}) = s^{-b} F_C(t,s;\vec{0}) = s^{-b} f_C(t/s)$ has the explicit scaling function\footnote{
Recast the integral $\int_0^1\!\D u\: u^{d/2-1} \left[ (y+1)-2u\right]^{-d/2} 
= \frac{2}{d} (1+y)^{-d/2} {}_2F_{1}\left( d/2,d/2;1+d/2;2/(1+y)\right)$\\ $=(y-1)^{-d/2}\, B_{2/(1+y)}\left(d/2,1-d/2\right)$ 
as a hypergeometric or incomplete Beta function \cite[eqs.(15.3.1,26.5.23)]{Abra65}.}
\BEA
f_C(y) &=& \frac{4T_c(d)}{d(4\pi)^{d/2}} \frac{y^{(2-d)/4}}{(1+y)^{d/2}} \,{}_2F_{1}\left( \frac{d}{2},\frac{d}{2};1+\frac{d}{2};\frac{2}{1+y}\right) 
\sim y^{-(3d-2)/4} \mbox{\rm ~;~~ if $0<d<2$} 
\nonumber \\
f_C(y) &=& \frac{4T_c(d)}{d(4\pi)^{d/2}} (1+y)^{-d/2} \,{}_2F_{1}\left( \frac{d}{2},\frac{d}{2};1+\frac{d}{2};\frac{2}{1+y}\right)  
\sim y^{-d/2} \hspace{0.6truecm}\mbox{\rm ~;~~ if $2<d$}
\label{4.10}
\EEA
along with the asymptotics for $y$ large. The values of the exponents $b$ and $\lambda_C$ are listed in table~\ref{tab1}. 

In analogy with the known results of the kinetic spherical model of magnets \cite{Hase06,Ebbi08}, 
the behaviour {\em at} the upper critical dimension
$d=d^*=2$ of the two-time observables $R(t,s;\vec{r})$ and $C(t,s;\vec{r})$ 
can be obtained by formally taking the limit $d\to 2$ in 
eqs.~(\ref{4.7}) and (\ref{4.10}). Hence {\em no} logarithmic
modifications of the leading scaling behaviour arise, but there are additive logarithmic corrections to scaling. 
This is different for single-time quantities
like the interface width $w(t)$, of which the leading long-time behaviour is
\BEA
w^2(t) &=& \frac{2\pi\, T_c(d)}{(8\pi)^{d/2}\sin(\pi d/2)} \, t^{1-d/2} \mbox{\rm ~;~~ if $0<d<2$} \nonumber \\
w^2(t) &\sim& 2 T_c \ln t \hspace{2.96truecm}\mbox{\rm ~;~~ if $d=2$} \label{4.11} \\
w^2(t) &=& 2T_c(d)\, {\cal F}_d + {\rm O}(t^{1-d/2}) \hspace{0.37truecm}\mbox{\rm ~;~~ if $d>2$} \nonumber
\EEA
with the constant ${\cal F}_d := \int_0^{\infty} \!\D\tau\: \left( e^{-4\tau} I_0(4\tau)\right)^d$, which diverges for 
$d\searrow 2$.\footnote{${\cal F}_3=\pi^{-2}\left(18+12\sqrt{2\,}-10\sqrt{3\,}-7\sqrt{6\,}\:\right)
{\bf K}^2\left[(2-\sqrt{3})(\sqrt{3}-\sqrt{2})\right]\simeq 0.1263655\ldots$, where
${\bf K}(k)$ is the complete elliptic integral \cite[eq.(2.15.21.2)]{Prudnikov2}.} 
Hence, the interface width
$w(t)\sim \sqrt{\ln t\,}$ growths logarithmically  in $2D$, see table~\ref{tab1}, 
in agreement also with experimental findings \cite{Salv96}. 
The values of exponent $\beta$, listed in table~\ref{tab1}, 
are the same as for the {\sc ew}-universality class.

Inspection of the values of the various exponents collected in table~\ref{tab1} shows that
\begin{enumerate}
\item the stationary exponents $z$, $\beta$, 
and consequently also the roughness exponent $\alpha=z\beta$, of the Arcetri model are in all dimensions the same 
as those of the Edwards-Wilkinson model. This might have been anticipated, 
since the equation of motion (\ref{3.2}) is still linear -- and also
reflects properties of the magnetic spherical model, where the equilibrium critical exponent $\beta=\demi$ of the order parameter and 
$\eta=0$ of the spin-spin correlator keep their mean-field values \cite{Berl52}. 
\item the non-equilibrium autocorrelation/autoresponse exponents  
$\lambda_C=\lambda_R$ depend on the dimensionality in a non-trivial way. Only for $d\geq d^*=2$, their values are the same as in the 
{\sc ew}-universality class. 

In this respect, the dynamics of the Arcetri model is quite similar 
to what is found in the non-equilibrium critical dynamics of {\em simple magnets}
with a non-conserved order parameter and disordered initial states. 
For such systems, a long-standing result of field-theory \cite{Jans89,Cala05,Taeu14} 
asserts that the true non-equilibrium exponents $\lambda_C,\lambda_R$ should be independent of the stationary exponents (since an
independent renormalisation is required for their calculation \cite{Jans89}).
The r\^ole of the stationary exponents is taken here by $z,\alpha,\beta$ and the exponents $a,b$ related to them. 
Hence the Arcetri model with $d<2$ can be considered as an explicit example of this general fact. 

We also recall that in the {\sc kpz} class with $d<2$, a Ward identity prevents this additional renormalisation, to all orders in perturbation
theory \cite{Krec97}, which is a qualitatively different situation. We are not aware of any estimate of $\lambda_C$ 
from a non-perturbative renormalisation-group study in the {\sc kpz} universality class, for $d\geq 2$. 
\item for the {\sc kpz}-universality class, there is a 
conjecture,\footnote{For the {\sc kpz} class in $d<2$ dimensions, $\lambda_C=d$ has been  proven 
to all orders in perturbation theory \cite{Krec97}.} for a flat interface, that 
$\lambda_C\stackrel{!}{=}d$ \cite{Kall99}; this also happens to be satisfied in the {\sc ew}-model. 
However, this conjecture does not extend to the Arcetri model, since $\lambda_C=\lambda_R\ne d$ for $d<2$.  
\item in the Arcetri model, autocorrelation and autoresponse exponents are always equal. 
Is the available numerical evidence for $\lambda_C\ne \lambda_R$ in the $2D$ {\sc kpz}-model \cite{Odor14} the final word~? 
\end{enumerate}  
\begin{figure}[tb]
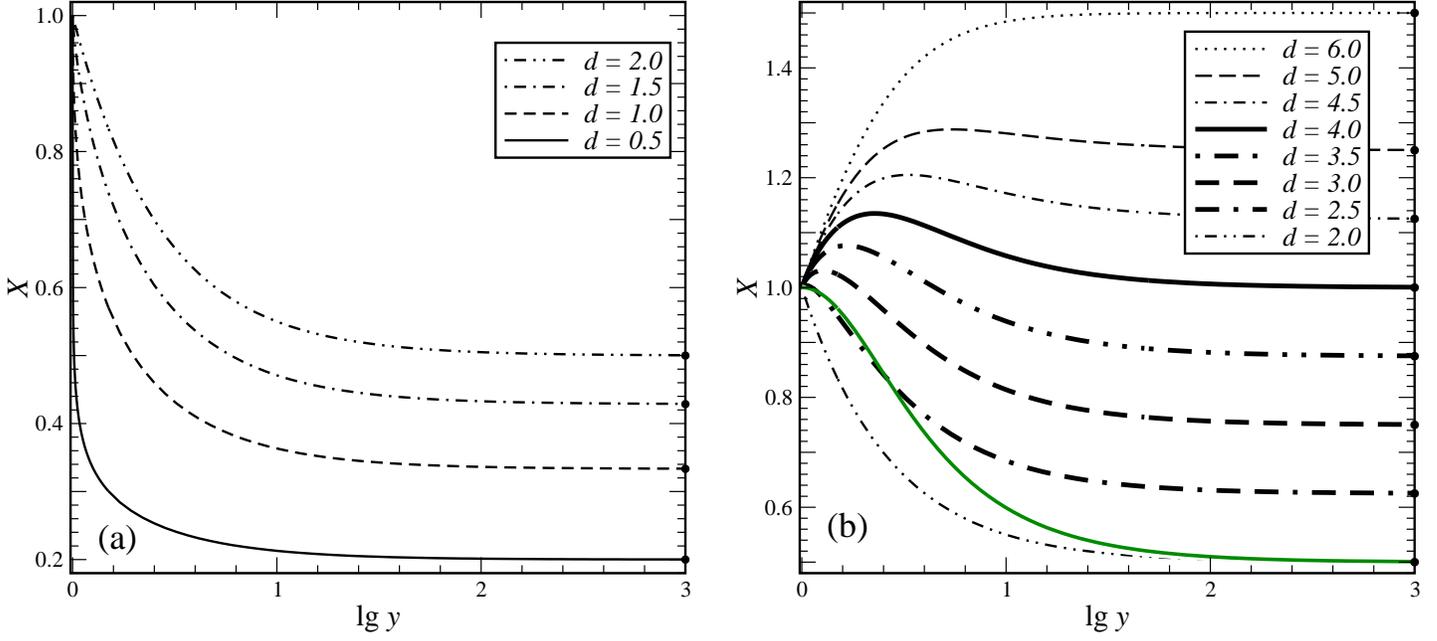

\centerline{\psfig{figure=durang5_arcetri_fig2a,width=3.6in,clip=} ~~
\psfig{figure=durang5_arcetri_fig2b,width=3.6in,clip=} }
\caption[fig2]{Fluctuation-dissipation ratio $X(y)$ over against $y=t/s$ of the critical Arcetri model for (a) $d\leq 2$ and 
(b) $d\geq 2$. The small circles on the right axis indicate the limit ratio $X_{\infty}$, eq. (\ref{4.13}). 
In (b), the solid green line gives $X(y)$ for the
{\sc ew}-model in $d=4$ dimensions.  \label{fig2}
}
\end{figure}
Remarkably, {\em even for $d>2$, the Arcetri and the {\sc ew} models are 
in different universality classes, in spite of all their exponents being equal}. 
In order to see this, recall that because of $a=b$, one can define the fluctuation-dissipation ratio (FDR) \cite{Cugl94} 
(for reviews, see \cite{Cris03,Cala05,Marc08,Leuz09,Cugl11})
\BEQ\label{4.12}
X(t,s) := T R(t,s) \left( \frac{\partial C(t,s)}{\partial s}\right)^{-1} = X\left(\frac{t}{s}\right)
\EEQ
where the last relation holds in the scaling limit. In magnetic systems or spin glasses, one uses the value of $X(y)-1$ 
as a measure of the distance with the respect to an equilibrium state. According to the Godr\`eche-Luck conjecture \cite{Godr00b}, 
the limit FDR $X_{\infty} = \lim_{y\to\infty} X(y)$ should be an universal number. 
For the critical Arcetri model, the explicit results (\ref{4.7}),(\ref{4.10}) lead to
\BEQ\label{4.13}
X_{\infty} = \left\{ \begin{array}{ll} d/(d+2) & \mbox{\rm ~;~~ if $0<d<2$} \\
                                       d/4     & \mbox{\rm ~;~~ if $d\geq 2$}
\end{array} \right.
\EEQ
Eq.~(\ref{4.13}) is distinct from the well-known value $X_{\infty}^{(\mbox{\footnotesize\sc ew})}=\demi$ of the {\sc ew}-model, 
valid in any dimension \cite{Roet06}.\footnote{The slope autocorrelator $A(t,s)$ and autoresponse $Q(t,s)$ of the $d$-dimensional Arcetri model
have the same scaling behaviour as the magnetic autocorrelator and autoresponse in the $(d+2)$-dimensional spherical model, see sec.~3.4. For
$d<2$, the limit `slope FDR' $X^{\footnotesize\rm SM}_{\infty}(d+2) = 1 - 2/(d+2) = d/(d+2)$ \cite{Godr00b} 
agrees with the {\em height} FDR (\ref{4.13}) of the $d$-dimensional Arcetri model.} 
To illustrate this further, we show in figure~\ref{fig2} the scaling function $X(y)$ as a function of $y=t/s$, 
for $d$ below (figure~\ref{fig2}a) and above (figure~\ref{fig2}b) the upper critical dimension $d^*=2$. 
Starting from the value $X(1)=1$ at equal times, 
the limit ratio $X_{\infty}$ is continuously approached for an increasing temporal separation $y=t/s\to \infty$. Qualitatively, this
approach is monotonous for dimensions $d\leq 2$, but for $d>2$ but not too large, 
$X(y)$ goes through a maximum before approaching its limit value. For sufficiently small
dimensions, this approach is from above, but when $d$ becomes large enough, the maximum of $X(y)$ disappears and the limit value
$X_{\infty}$ is approached monotonously from below. 
Curiously, for $d=4$ one has $X_{\infty}=1$. 

For $d=2$, the FDR of the {\sc ew}-model is identical to the
Arcetri model and the green solid line in figure~\ref{fig2}b gives the FDR of the {\sc ew}-model for $d=4$. 
For all dimensions $2<d<4$, the FDRs of the {\sc ew}-model fall
between these two curves, which are quite distinct from those of the Arcetri model. 

\begin{figure}[tb]
\centerline{\psfig{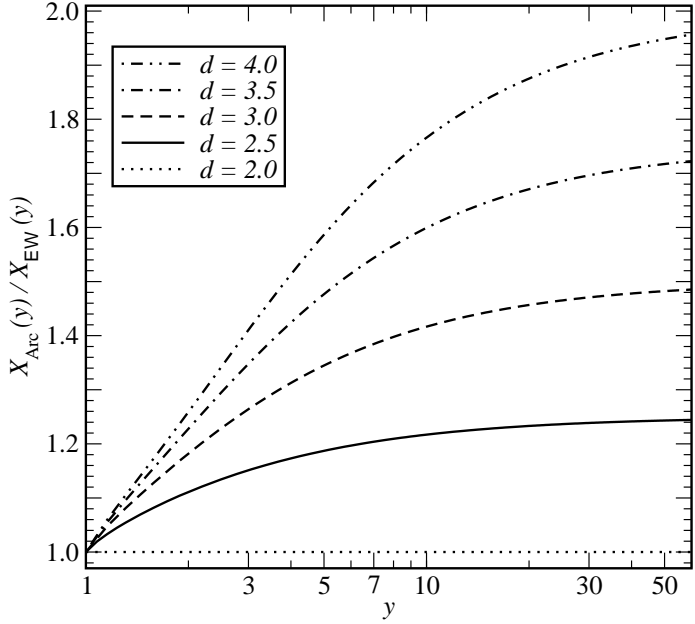} }
\caption[fig3]{Ratio of the fluctuation-dissipation ratio $X_{\rm Arc}(y)$ for the Arcetri model and of $X_{\rm EW}(y)$ for the
{\sc ew}-model, as a function of $y=t/s$, for several dimensions $2\leq d\leq 4$. \label{fig3}
}
\end{figure}

The distinction between the Arcetri and {\sc ew}-models is further illustrated in figure~\ref{fig3}, 
where the ratio of the two fluctuation-dissipation ratios is plotted. With an increasing separation $y=t/s>1$ of the time-scales
the distinction of the two models becomes progressively more clear. Only for $d=2$ the two functions $X(y)$ are identical, hence the
two models are indeed identical in $2D$.\footnote{Therefore, the experiments described in \cite{Salv96} apply to both.}  

Hence the Arcetri model with $d\geq 2$ illustrates the usefulness of the fluctuation-dissipation ratio as a diagnostic tool 
for a fine distinction of non-equilibrium
universality classes. It also points to the interest of measuring experimentally both two-time correlators 
and responses in growing interfaces (long-range interactions generically reduce the value of $d^*$). 

\subsection{Relationship with the bosonic pair-contact process} 

Surprisingly, it turns out that {\em for $2<d<4$, 
the non-equilibrium relaxation properties of the critical Arcetri model are related to those of a different stochastic
process, the so-called (multi-)critical {\em {\bf b}osonic {\bf p}air-{\bf c}ontact {\bf p}rocess with {\bf d}iffusion}} ({\sc bpcpd}) 
\cite{Houc02,Paes04,Baum05}, see \cite{Howa97} for a field-theoretic description. In this model, each site of a 
$d$-dimensional hypercubic lattice can be occupied by an arbitrary number $n\in\mathbb{N}$ of particles of a single species $A$. 
Single particles may hop to a nearest-neighbour site, with a diffusion rate $D$. 
On the same site, pairs of particles may react, according to the schemes 
$2A \to (2+k)A$ or $2A\to (2-k)A$, where $k$ is either one or two. 
The universal long-time behaviour of the model does not depend on the value of $k$. 
The two reaction rates are chosen equal $\Gamma[2A\to (2+k)A] = \Gamma[2A\to (2-k)A] = \mu$. 
The model's behaviour is determined by the value of the control parameter $\alpha := k^2 \mu/D$. 
There is a critical value, $\alpha=\alpha_C$, with\footnote{Using ${\cal F}_3$ from (\ref{4.11}), 
$\alpha_C = (2 {\cal F}_3)^{-1}\simeq 3.956776\ldots$ for $d=3$.}
\BEQ
\frac{1}{\alpha_C} = 2 \int_0^{\infty} \!\D u\: \left( e^{-4u} I_0(4u)\right)^d
\EEQ
and for $d>2$, one has $\alpha_C>0$. For $\alpha>\alpha_C$, all particles cluster on a few sites only \cite{Paes04,Baum05,Henk10} 
while for $\alpha<\alpha_C$, the system evolves
towards a spatially homogeneous stationary state. We are interested in the multi-critical point at $\alpha=\alpha_C>0$, 
where this `{\em clustering transition}' occurs.
For $d\leq 2$, the multi-critical point in the {\sc bpcpd} does not exist.

The dynamics of the model is described in terms of the bosonic 
particle-annihilation operator $a(t,\vec{r})$ such that the particle-number operator is
$n(t,\vec{r})=a^{\dag}(t,\vec{r})a(t,\vec{r})$. For any value of $\alpha$, 
and an initial state with uncorrelated particles of a mean number density $\rho_0$,
the average particle-number $\left\langle n(t,\vec{r})\right\rangle=\left\langle a(t,\vec{r})\right\rangle = \rho_0$ 
is constant \cite{Paes04,Baum05}. The two-time density-density correlator can be expressed as \cite{Henk10}
\BEQ
\left\langle n(t,\vec{0}) n(s,\vec{r}) \right\rangle - \rho_0^2 = \bar{C}(t,s;\vec{r}) + \bar{R}(t,s;\vec{r})\rho_0
\EEQ
with the following definition of the two-time correlators and responses \cite{Baum05}
\BEQ
\bar{C}(t,s;\vec{r}) = \left\langle a^{\dag}(t,\vec{0}) a(s,\vec{r}) \right\rangle - \rho_0^2 
\;\; , \;\;
\bar{R}(t,s;\vec{r}) = \left. \frac{\delta \left\langle a(t,\vec{0})\right\rangle}{\delta j(s,\vec{r})}\right|_{j=0}
\EEQ
where $j$ denotes the particle-creation rate conjugate to the particle number. 
The usual scaling behaviour (\ref{1.2}),(\ref{1.3}) has been confirmed, for 
all $\alpha\leq \alpha_C$. Precisely {\em at} the clustering transition $\alpha=\alpha_C$, 
and for $2<d<4$ dimensions, the explicit scaling behaviour reads, for $s\to\infty$ \cite{Baum05}
\BEQ\label{4.15}
\bar{C}(ys,s;\vec{0}) = c_0 \rho_0^2 \, (y+1)^{-d/2} {}_2F_{1}\left(\frac{d}{2},\frac{d}{2};1+\frac{d}{2};\frac{2}{1+y}\right)
\;\; , \;\; 
\bar{R}(ys,s;\vec{0}) = s^{-d/2} r_0 (y-1)^{-d/2}
\EEQ
and where $c_0, r_0$ are known normalisation constants. Hence, $\bar{C}(t,s;\vec{0})$ can be considered as the scaling limit 
of the connected density-density autocorrelator in the {\sc bpcpd}. 
For $2<d<4$, the shape of the scaling functions is identical to the ones of the Arcetri model
given in eqs. (\ref{4.10}),(\ref{4.7}). However, although $a=d/2-1$ in both models, the values of the ageing exponents are distinct, 
namely $b=0$ for the {\sc bpcpd} and $b=d/2-1$ for the Arcetri model. Since $a\ne b$, the fluctuation-dissipation ratio (\ref{4.12}) cannot
be defined in  the multi-critical {\sc bpcpd}. 

This is a further example which illustrates that the values of the non-equilibrium exponents $\lambda_{C,R}$ are independent from the
stationary exponents $\alpha,\beta,z$, analogously to non-equilibrium
field theory of critical magnets \cite{Jans89,Cala05,Taeu14}, but different from the {\sc kpz} universality class \cite{Krec97,Taeu14}. 
However, this independence is illustrated in the Arcetri model for $2<d<4$ in a new way. Recall that for $d<2$, all the exponents 
$\alpha,\beta,z$ and $a,b$ are found to be the same in the {\sc ew} and Arcetri universality classes, yet the values of $\lambda_C=\lambda_R$ 
are different. Here, when $2<d<4$, the values of the non-equilibrium exponents $\lambda_C=\lambda_R=d$, as as well the ageing exponent $a=d/2-1$,
are the same. However, the ageing exponent $b=0$ in the {\sc bpcpd} and $b=d/2-1$ in the Arcetri model are different. 

Finally, for $d>4$, the shape of $f_C(y)$ in the  multi-critical {\sc bpcpd} is different from (\ref{4.15}), 
although $\lambda_C=d$, and $b=d/2-2$ \cite{Baum05}. 
This gives a different example, but of the same kind, as found for $d<4$. 


We emphasise that although the value of the autocorrelation exponent $\lambda_C=d$, 
the two-time correlation functions are {\em distinct} from those of the {\sc ew}-model, 
where $C(t,s)=s^{-b} f_C^{\rm\footnotesize (EW)}(t/s)$ with $b=d/2-1$ and
$f_C^{\rm\footnotesize (EW)}(y) = \frac{2 T}{2-d}(4\pi)^{-d/2} \left[ (y+1)^{1-d/2} - |y-1|^{1-d/2}\right]$ \cite{Roet06}. 
This rather co\"{\i}ncides with the shape of 
$f_C(y)$ in the critical {\sc bpcpd} with $\alpha<\alpha_C$ \cite{Baum05}.

\subsection{Ageing for $T<T_c$}

In order to discuss the behaviour below the critical point, when $T<T_c$ and the noise is thus relatively weak, it is useful
to introduce the auxiliary variable $m^2 := 1-T/T_c$. 

The two-time response function again takes the scaling form (\ref{4.6}), where the scaling function $f_R(y)$ of the autoresponse now becomes
\BEQ \label{4.16}
f_R(y) = (4\pi)^{-d/2} y^{(d+2)/4} (y-1)^{-d/2} 
\EEQ
for all dimensions $d>0$. Similarly, the two-time autocorrelator becomes 
(with $t=ys>s$ assumed; the first correction is valid for $d<2$, see appendix~A)
\BEQ \label{4.17}
C(t,s) =  2^{d+2} m^2 s\, y^{(2+d)/4} (y+1)^{-d/2} + {\rm O}\left( T s^{1-d/2}\right)
\EEQ
which has the expected scaling form $C(t,s) = s^{-b} f_C(t/s)$ with the explicit scaling function
\BEQ
f_C(y) = 2^{d+2} m^2 \: y^{(2+d)/4} (y+1)^{-d/2}
\EEQ
The non-equilibrium exponents can be read off and are listed in table~\ref{tab1}. 
The extension to time-space correlations $C(t,s;\vec{r})$ is straightforward. 
The interface width is obtained as the special case $w^2(t)=C(t,t)$ and reads (the first correction is valid for $d<2$)
\BEQ\label{4.19}
w^2(t) = 4 m^2 t + {\rm O}\left( T t^{1-d/2}\right)
\EEQ
For correlators and the interface width, the long-time behaviour only depends on the ratio $T/T_c$ 
and the short-ranged nature of the initial correlations (parametrised by $H_1$). 
The `thermal' convolution term, which dominated the critical case, merely gives rise to a correction to scaling. 
This leading result $w^2(t) \simeq 4m^2 t$, valid for $t$ large enough, is independent of $d$, hence the growth exponent $\beta=\demi$.

\subsection{Global persistence}

A different class of observables are first-passage probabilities, such as persistence probabilities, see \cite{Bray13} for a recent detailed review. 
The global persistence probability $P_{\rm G}(t)$ is the probability
that a fluctuation $\delta h_{\rm G}(t)$ of the {\em global} height 
$h_{\rm G}(t) \sim \int_{\mathbb{R}^d} \D\vec{r}\,  h(t,\vec{r})\sim \wht{h}(t,\vec{0})$ did never change its sign
between the initial instant $t=0$ and time $t$. For large times, one expects a power-law decay $P_{\rm G}(t)\sim t^{-\theta_{\rm G}}$, where
$\theta_{\rm G}$ is the {\em global persistence exponent}. In the infinite-size limit, the central limit theorem implies that the global height should  
be a gaussian variable \cite{Maju96,Bray13}. Remarkably, it can be shown that then the model's underlying stochastic process is markovian if and only if
the normalised global autocorrelator
\BEQ \label{4.22}
\wht{N}(t,s) := \frac{\wht{C}_{\vec{0}}(t,s)}{\sqrt{ \wht{C}_{\vec{0}}(t,t)\,\wht{C}_{\vec{0}}(s,s)\,}\,} = \left( \frac{s}{t}\right)^{\mu}
\EEQ
(where $\wht{C}_{\vec{0}}(t,s)$ is the two-time autocorrelator in momentum space) 
takes in the scaling limit, $t,s\to\infty$ with $y=t/s$ fixed, a simple power-law form \cite{Maju96}. Then $\theta_{\rm G}=\mu$, and in addition, 
a scaling relation relating $\theta_{\rm G}$ and the
autocorrelation exponent $\lambda_C$ can be derived, for quenches either to $T=T_c$ \cite{Maju96} or to $T<T_c$ \cite{Cuei99,Henk09b,Bray13}. 
Adapting this to the exponent notation (\ref{1.2}),(\ref{1.3}) used for growing interfaces, this `markovian' scaling relation reads
\BEQ \label{4.23}
\theta_{\rm G} = \frac{1}{z} \left( \lambda_C - \frac{d}{2} - \frac{z b}{2} \right) \geq 0
\EEQ
where the bound follows from a Yeung-Rao-Desai inequality \cite{Yeun96}, see appendix~B. 
However, in many non-equilibrium systems, quenched to either $T=T_c$ or $T<T_c$, those scaling relations
turn out to be broken (albeit by numerically small amounts) so that one 
should conclude that the underlying stochastic processes cannot be markovian at
sufficiently large times, see \cite{Bray13,Henk10} and references therein for explicit examples. Practically, deviations of 
$\wht{N}(t,s)$ from a pure power law may be more easy and more reliable to detect than tiny deviations from (\ref{4.23}) \cite{Henk09b}.

Applying this general method to the Arcetri model, with (\ref{3.13}) the global height autocorrelator is
$\wht{C}_{\vec{0}}(t,s) = \left( H_1 +2\Gamma T\int_0^s\!\D\tau\, g(\tau)\right)/\sqrt{g(t)g(s)\,}$, for $t\geq s$. 
Recalling the leading long-time behaviour of $g(t)$ from appendix~A, we have
\BEQ
\wht{N}(t,s) = 
\sqrt{\frac{H_1 +2\Gamma T\int_{0}^{s} \!\D\tau\, g(\tau)}{H_1 +2\Gamma T\int_{0}^{t} \!\D\tau\, g(\tau)}}
\stackrel{t,s\to\infty}{=} \left( \frac{s}{t}\right)^{\mu} \;\; ,\;\;
\mu=\theta_{\rm G}= \left\{ \begin{array}{ll} 
0 & \mbox{\rm ~;~~ if $T<T_c$ and $d>0$} \\
d/4 & \mbox{\rm ~;~~ if $T=T_c$ and $0<d<2$} \\
1/2 & \mbox{\rm ~;~~ if $T=T_c$ and $d>2$} 
\end{array} \right.
\EEQ
As expected, the verification of the simple power-law (\ref{4.22}) 
confirms the Markov property, hence the identification $\mu=\theta_{\rm G}$ is admissible.

\section{Discussion and perspectives}

{\it Concluendum est:} this work explores the properties of an analogue of the familiar spherical model of ferromagnets \cite{Berl52} 
for the description of growing interfaces. 
In the scheme presented here, we admitted that the correspondence should be made between the local {\em slopes} of the interface and the spherical spins. 
It then becomes relatively straightforward to write down and solve the Langevin equations of motion, since essentially all techniques can be 
borrowed from the kinetic spherical model \cite{Ronc78,Cugl95,Godr00b}. 
In particular, we have analysed the long-time behaviour of the Arcetri model, defined in section~2. 
Our findings are as follows:
\begin{enumerate}
\item the parameter $T$, named a `temperature' by analogy with the kinetic spherical model, 
and defined through the noise correlator (\ref{2.2}), admits for all dimensions $d>0$ a critical value $T_c=T_c(d)$, given by eq. (\ref{4.4}). 
Qualitatively, the Arcetri model is in this respect more analogous to magnetic systems and qualitatively different from more usual interface growth
universality classes, such as described by the {\sc kpz} or  {\sc ew} universality classes. 

At criticality $T=T_c(d)$, starting from a flat substrate, the interface becomes rough for $d\leq 2$ and remains smooth for $d>2$. 
For $T<T_c(d)$, the interface always becomes rough and for $T>T_c(d)$, even an initially rough interface becomes smooth in a finite time. 
\item for $T>T_c(d)$, the model's relaxation behaviour is rapid, time-translation-invariant and governed by a finite relaxation time.
\item in  contrast, for $T\leq T_c$, the relaxation is slow, time-translation-invariance is broken and the two-time observables obey dynamical
scaling (\ref{1.2}),(\ref{1.3}) for large times. Hence the conditions for {\em physical ageing}, as defined in \cite{Henk10}, are obeyed. 

Such a physical ageing behaviour has already been found in more standard models of growing interfaces, 
especially in the {\sc kpz}- and  {\sc ew}-models. 
In table~\ref{tab1}, the values of the relevant exponents are listed. 
\item the critical Arcetri model has an upper critical dimension $d^*=2$. 

If $d=d^*=2$, no logarithmic modifications of the dynamical scaling arise for the two-time observables, but a logarithmic behaviour
is seen for single-time quantities such as the interface width. The $2D$ Arcetri and {\sc ew}-models are identical.
\item in many respects, if one chooses $T=T_c$, the resulting behaviour of the critical Arcetri model is qualitatively quite analogous to what has been
found in other universality classes. However, some differences can be seen as well:
\begin{enumerate}
\item the stationary exponents $z,\beta,\alpha$ (as well as the exponents $a,b$ which depend on them), are the same for the {\sc ew}-model and the
Arcetri model. However, for $d<2$, the non-equilibrium exponents $\lambda_C$ and $\lambda_R$ of both models are different. 

In this respect, the critical Arcetri model behaves analogously to the non-equilibrium critical dynamics of simple non-conserved magnets and is
qualitatively different from growth models in the {\sc kpz} universality class \cite{Krec97}. 
It provides an explicit example where, for disordered initial states, the non-equilibrium
exponents $\lambda_{C,R}$ are independent (i.e. not related by a scaling relation) from the stationary exponents. This is a long-standing
prediction from the non-equilibrium critical dynamics of simple non-conserved magnets \cite{Jans89,Cala05,Taeu14}. 
\item in the Arcetri model, one has always $\lambda_C=\lambda_R$. This matches the exact results of the {\sc ew}-model and the $1D$ {\sc kpz}-model.
Curiously, the available numerical data indicate the contrary for the $2D$ {\sc kpz}-model \cite{Odor14}. 
\item the conjecture $\lambda_C\stackrel{!}{=}d$ \cite{Krec97,Kall99}, formulated for the {\sc kpz} equation, and observed to hold in the {\sc ew}-model
as well, does not extend to the Arcetri model when $d<2$. 

The available numerical data for the $2D$ {\sc kpz}-model \cite{Odor14,Halp14} 
do not agree with it either. We remark that for the $1D$ {\sc kpz} equation, 
the shape of the two-time response function could only be explained using 
the logarithmic extension of local scale-invariance, which implies logarithmic
terms in the scaling function \cite{Henk12,Henk13}. {\em If} such an observation 
could be extended to the $2D$ case as well: could  
un-recognised logarithmic contributions have modified the effective values of $\lambda_{C}$ or $\lambda_{R}$~?  
\end{enumerate}
\item for dimensions $2<d<4$, the two-time response function, 
and the shape of the scaling functions of the two-time correlator (up to normalisation),
are the same as at the multi-critical point in the so-called 
{\it bosonic pair-contact process with diffusion} ({\sc bpcpd}) \cite{Baum05}. 
However, although the asymptotic exponent
has the same value $\lambda_C=d$, the shape of the scaling function 
of the two-time correlator of the {\sc ew}-model is distinct from the one of
the Arcetri model for $d>2$, although the two-time responses $R(t,s;\vec{r})$ and the exponent $b=d/2-1$ agree. 
\item an universal method to distinguish the Arcetri and {\sc ew}-models for 
$d>2$ uses the fluctuation-dissipation ratio (\ref{4.12}) \cite{Cugl94}, 
in  particular the universal limit FDR $X_{\infty}$ \cite{Godr00b} 
has different values in  the two universality classes.\footnote{The distinction 
between the two models for $d\ne 2$ should be related to the spherical constraint (\ref{2.4}), 
which is incompatible with a flat surface $h_n=\mbox{\rm cste.}$.} 

Hence, for $d>2$, we have three distinct models, with the same values of the exponents $a$, 
and $\lambda_R=\lambda_C$, but which are distinguished
as follows: (i) between the Arcetri and {\sc bpcpd}-classes, the exponent $b$ has different values. 
(ii) between the Arcetri and {\sc ew}-classes, the FDR and in particular the limit $X_{\infty}$ are different. 
This illustrates once more the independence of the exponent $\lambda_C$ from the other ones. 
\item the relationship with the kinetic spherical model in $d+2$ dimensions (see section~3.4), 
relating local slopes to local magnetisations, 
allows to draw a clear physical picture on how the long-time relaxation is going on: 
for $T<T_c$, the Arcetri model undergoes a coarsening, such that increasingly larger patches of the interface 
(of typical size $L(t)\sim t^{1/2}$ for all $d$ and $T\leq T_c$ \cite{Ronc78,Ebbi08}) 
have constant slope, leading to a saw-tooth pattern. For $T=T_c$, the interface slopes 
are correlated over increasing distances $L(t)$, such that
the interface itself should become a fractal. 
\end{enumerate}

The spherical model is easily adapted and generalised, for instance to long-range initial conditions \cite{Pico02,Dutt08}, 
long-range interactions \cite{Cann01,Baum07,Dutt08}, external fields \cite{Paes03},
conserved order parameters \cite{Coni94,Cann01,Sire04,Baum07b}, frustrations and/or disorder \cite{Cugl95}, 
external drives \cite{Hase12,Godr13} and so on. 
Several of those modifications could be of interest to experimentalists. 
It should be possible to generalise the Arcetri model in these directions, 
to work out the consequences on its ageing behaviour and explore
relationships with long-range particle-reaction models \cite{Dura09}. 

\textcolor{black}{Another important question will be if the Arcetri model can be understood as a $n\to\infty$ limit of a suitable 
$n$-component generalisation of the {\sc kpz}-equation, following the lines outlined in \cite{Dohe94}. We hope to come back to this elsewhere.}

Finally, the model analysed here is but one way to write down a `spherical analogue' 
of the {\sc kpz} universality class. We used here
the representation in terms of the Burgers equation
\BEQ \label{5.1}
\partial_t u = \nu \partial_x^2 u + \mu u \partial_x u + \partial_x \eta ~~\mapsto~~ 
\partial_t u = \nu \partial_x^2 u + \mathfrak{z}(t) u + \partial_x \eta
\EEQ
where $\mathfrak{z}(t) \sim \left\langle \partial_x u\right\rangle~\sim 1/t$ (which follows from $g(t)\sim t^{\digamma}$) 
might be seen as some kind of `averaged curvature' of the interface. 
\textcolor{black}{However, the choice considered here was also motivated by its mathematical simplicity. 
It might be thought that a more faithful representation of the structure of the non-linear terms could be achieved in terms of a replacement}
\BEQ \label{5.2}
\partial_t u = \nu \partial_x^2 u + \mu u \partial_x u+ \partial_x \eta ~~\mapsto~~ 
\partial_t u = \nu \partial_x^2 u + \mathfrak{z}(t) \partial_x u + \partial_x \eta
\EEQ
where $\mathfrak{z}(t) \sim \left\langle u\right\rangle$ 
might be viewed as some kind of `averaged slope'. In these two cases, \textcolor{black}{the Lagrange multiplier} $\mathfrak{z}(t)$ is to be
found from a spherical constraint $\sum_x \langle u^2\rangle  = {\cal N}$. 
Finally, we might have started directly from the {\sc kpz} equation
\BEQ \label{5.3}
\partial_t h = \nu \partial_x^2 h + \demi \mu \left( \partial_x h\right)^2 + \eta ~~\mapsto~~ 
\partial_t h = \nu \partial_x^2 h + \mathfrak{z}(t) \partial_x h +\eta
\EEQ
where $\mathfrak{z}(t) \sim \left\langle \partial_x h\right\rangle$ might again be interpreted 
as an `averaged slope' and will be found from a
constraint $\sum_x \langle\left( \partial_x h\right)^2\rangle = {\cal N}$. 
\textcolor{black}{While eq.} (\ref{5.1}) 
has been studied here, 
\textcolor{black}{work on the other two models, as defined in eqs.~(\ref{5.2}),(\ref{5.3}) is in progress. The analysis of their
sphericl constraints presents new features not seen in the system studied in this work  and apparently leads
to new types of behaviour. A sequel paper will take up the analysis of these distinct universality classes.}  

\newpage
%
%
\appsection{A}{Computation of the long-time behaviour}

The explicit long-time behaviour of the spherical constraint parameter $g(t)$, for $T\leq T_c$ 
and the consequences for the interface width $w(t)$ and the
two-time responses and correlators, stated in section~4, will be derived. 

Since $\lap{g}(p)$ is given by (\ref{4.1}), we use 
standard Tauberian theorems \cite[ch. XIII.5]{Fell71}, in order to obtain the 
long-time behaviour of $g(t)$ from the $p\to 0$ behaviour of $\lap{g}(p)$.
To do so explicitly, two methods have been used: (i) one may consider $g(t), f(t)$ as regular functions, which are found by
inverse Laplace transformation. These must be completed by certain sum rules, 
needed for the computation of the correlators, see e.g. \cite{Cugl95,Godr00b,Cann01,Hase06}. 
(ii) perhaps more straightforwardly, accept that $g(t),f(t)$ may have singular terms described by the Delta function 
and its derivatives, which are formally obtained by inverting the first few orders of $\lap{g}(p)$ \cite{Pico02,Ebbi08}, see below. 
Although these singular terms do not contribute to the responses, 
they are important when inserted into the integrals which give the width or the correlators, since the singular terms in $g(\tau)$ 
will reproduce the effect of the sum rules just mentioned. In conclusion, both methods lead to the same final result.

\subsection{Ageing at the critical point}

Here, we consider the long-time behaviour at the critical point $T=T_c(d)$, as given by (\ref{4.4}),(\ref{4.5}). 

In order to find $\lap{g}(p)$, we first expand $\lap{f}(p)=\int_0^{\infty} \!\D t\, e^{-pt} f(t)$, where $f(t)$ was given in (\ref{3.12}). 
In principle, such an expansion will contain terms analytic in $p$, i.e. $\sim p^n$ with $n\in\mathbb{N}$, 
and non-analytic contributions, i.e. $\sim p^{\alpha}$ with $\alpha\not\in\mathbb{N}$. The origin of both is clearly seen
by splitting the integral $\int_0^{\infty}\!\D t = \int_0^{\eta}\!\D t\, + \int_{\eta}^{\infty} \!\D t$ where $\eta$ is a cut-off to be sent to
infinity at the end. The first term can be formally expanded in $p$, 
as long as the corresponding coefficients converge, and gives the analytic part.
The non-analytic part is found by inserting the $t\to\infty$ asymptotics of $f(t)$ in the second term. 
To carry this out this, one sends first $p\to 0$ and only afterwards, one recovers the $\eta\to\infty$ limit.
This standard calculation, e.g. \cite{Joyc72,Luck85,Pico02}, leads to the expansions, with 
$A_d^{-1} = -\frac{2}{\pi} (8\pi)^{d/2}\sin\left(\frac{\pi d}{2}\right)\Gamma\left(\frac{d}{2}\right)$, $\lap{f}(0)=d/(2T_c(d))$  
and ${\lap{f}\,}'(0)=-\frac{d}{16}\int_0^{\infty}\!\D t\: e^{-dt} I_1(t) I_0(t)^{d-1}$
\BEQ \label{A1}
\lap{f}(p) \simeq \left\{ \begin{array}{ll} 
\frac{d}{2T_c(d)} + A_d\, p^{d/2}                      & \mbox{\rm ~;~~ if $0<d<2$} \\
\frac{d}{2T_c(d)} + \frac{1}{16\pi} p \left( \ln p + C_E -1\right) & \mbox{\rm ~;~~ if $d=2$} \\
\frac{d}{2T_c(d)} + {\lap{f}\,}'(0)\, p + A_d \, p^{d/2} & \mbox{\rm ~;~~ if $2<d<4$} 
\end{array} \right.
\EEQ
($C_E=0.5772\ldots$ is Euler's constant) and for $d>4$, a term ${\rm O}(p^2)$ 
appears which does not contribute to the leading scaling behaviour. 
Hence, with $G_d :=-\frac{H_1 d}{(2 T_c(d))^2}\frac{1}{{\lap{f}\,}'(0)}>0$ 
\BEQ
\lap{g}(p) \simeq \left\{ \begin{array}{ll} 
-\frac{d}{(2 T_c(d))^2}\frac{H_1}{A_d} \,p^{-d/2} - \frac{H_1}{2 T_c(d)} + {\rm o}(p) & \mbox{\rm ~;~~ if $0<d<2$} \\[0.16cm]
G_d\, p^{-1} - \frac{H_1}{2 T_c(d)} + \frac{H_1 A_d}{(2 T_c(d))^2}\frac{1}{({\lap{f}\,}'(0))^2}\, p^{d/2-1} 
+ {\rm O}(p^{d-2}) & \mbox{\rm ~;~~ if $2<d<4$} 
\end{array} \right.
\EEQ
and inverting this term-by-term gives, for $t\to\infty$ and with 
$g_d := -\frac{d}{(2 T_c(d))^2}\frac{H_1}{A_d} \frac{1}{\Gamma(d/2)}>0$ 
\BEQ \label{A3}
g(t) \simeq \left\{ \begin{array}{ll} 
g_d \,t^{d/2-1} - \frac{H_1}{2 T_c(d)}\, \delta(t)  & \mbox{\rm ~;~~ if $0<d<2$} \\[0.16cm]
G_d - \frac{H_1}{2 T_c(d)}\, \delta(t) + {\rm O}(t^{-d/2}) & \mbox{\rm ~;~~ if $d>2$} 
\end{array} \right.
\EEQ

In order to find the two-time response explicitly, it is enough to insert into (\ref{3.16}) 
the asymptotic form of $g(t)$ for $t\to\infty$, as derived above, to
simply drop any distributional terms\footnote{The singular terms in (\ref{A3}) do not contribute for $t$ large. Heuristically, recall:
$\delta(t) = \lim_{\lambda\to\infty}\sqrt{\lambda/\pi\,}\,e^{-\lambda t^2}$ vanishes for $t\ne 0$.} and to expand $F_{\vec{r}}(t)$ 
for large times as well.\footnote{The relevant Bessel asymptotic formula is
$I_{r}(t) \simeq (2\pi t)^{-1/2} e^{t-r^2/(2t)}(1+{\rm O}(t^{-1})$ for $t\to\infty$.} This immediately gives (\ref{4.6}),(\ref{4.7}).

Next, we analyse the interface width $w(t)$. Consider first the case $0<d<2$. From (\ref{3.24}), it follows
\BEA
w^2(t) &=& \frac{H_1}{g(t)} F_{\vec{0}}(2t) + \frac{2 T_c}{g(t)} \int_0^{t} \!\D\tau\: g(\tau) F_{\vec{0}}(2t-2\tau)
\nonumber \\
&\simeq& \frac{H_1}{g(t)} F_{\vec{0}}(2t) 
+ \frac{2 T_c}{g(t)} \int_0^{t} \!\D\tau\: \left[ - \frac{H_1}{2 T_c} \delta(\tau) - g_d \tau^{d/2-1} \right] F_{\vec{0}}(2t-2\tau)
\nonumber \\
&\simeq& 2 T_c t^{1-d/2} \int_0^{t} \!\D\tau\:\tau^{d/2-1} F_{\vec{0}}(2t-2\tau)
\nonumber \\
&=& 2 T_c t^{1-d/2} {\cal L}^{-1} \left( \left( {\cal L}\tau^{d/2-1}\right)(p)\left( {\cal L}F_{\vec{0}}(2\tau)\right)(p)\right)(t)
\nonumber \\
&\simeq& 2 T_c \frac{\Gamma(1-d/2)\Gamma(d/2)}{(8\pi)^{d/2}} \left({\cal L}^{-1} \frac{1}{p}\right)(t)\, t^{1-d/2}
\nonumber \\
&=& \frac{2\pi}{(8\pi)^{d/2}\sin(\pi d/2)} \, T_c(d) \, t^{1-d/2}
\label{A4}
\EEA
which gives the first line in (\ref{4.11}). In the second line in (\ref{A4}), 
we used in the integral the asymptotic form of $g(\tau)$, including both the
singular distributional as well as the leading regular term. 
Integrating the singular term, we see that it cancels against the other contribution to
$w^2(t)$. In the third line, we also inserted the asymptotic form of $g(t)$. 
As for the response before, the singular terms do not contribute for
$t\to\infty$. Furthermore, the non-universal amplitudes $g_d$ contained in $g(\tau)$ and $g(t)$ cancel. 
In the forth line, we recognise the remaining integral as a Laplace convolution which is treated via Laplace transforms. 
In the fifth line, we inserted the small-$p$ behaviour of the two Laplace transforms, namely 
$\left({\cal L}\tau^{d/2-1}\right)(p)=\Gamma(d/2) p^{-d/2}$ and 
\BD
\left( {\cal L}F_{\vec{0}}(2t)\right)(p) = \int_0^{\infty} \!\D t\: e^{-pt} \left( e^{-4t} I_0(4t)\right)^d 
\stackrel{p\to 0}{\simeq} \int_{\eta}^{\infty}\!\D t\: e^{-pt} (8\pi t)^{-d/2} + {\rm O}(1)
\stackrel{p\to 0}{\simeq}  \frac{\Gamma\left( 1- d/2\right)}{(8\pi)^{d/2}} p^{d/2-1}
\ED
In analogy with the computation above of $\lap{f}(p)$, the integral diverges for $d<2$ 
in the $p\to 0$ limit. Hence its leading contribution in this limit can be found by splitting the
integral $\int_0^{\infty} = \int_0^{\eta} + \int_{\eta}^{\infty}$ with a cut-off $\eta$. The contribution of the
first term remains finite for $p>0$ and $\eta<\infty$ and it is enough here to extract the leading, divergent, behaviour of the second term. 
Herein, the $\Gamma$-functions are defined via analytic continuation, if necessary \cite{Abra65}. On the other hand, for $d>2$, 
we first find the same cancellation of the $H_1$-dependent term and then, using (\ref{A3})
\BEQ
w^2(t)\simeq 2T_c \int_0^t \!\D\tau\, F_{\vec{0}}(2t-2\tau) \nonumber \\
= 2T_c  \underbrace{\int_0^{\infty} \!\D\tau\, F_{\vec{0}}(2\tau)}_{=:{\cal F}_d} - 2 T_c \int_t^{\infty} \!\D\tau\, F_{\vec{0}}(2\tau) 
= 2T_c {\cal F}_d + {\rm O}(t^{1-d/2})
\EEQ
since the amplitudes $G_d$ cancel. The last integral was estimated by using 
$F_{\vec{0}}(t) \sim t^{-d/2}$, for $t$ large enough. This gives the last line in (\ref{4.11}). 

For the two-time correlator, consider first the case $0<d<2$. We can use the same cancellation mechanism as before
for the temperature-independent term and then find, assuming $t=ys\geq s$
\BEA
C(t,s;\vec{r}) &=& \frac{H_1}{\sqrt{ g(t)g(s)\,}\,} F_{\vec{r}}(t+s) 
+ \frac{2T_c}{\sqrt{g(t) g(s)\,}\,} \int_0^s \!\D\tau\: g(\tau) F_{\vec{r}}(t+s-2\tau)
\nonumber \\
&\simeq& \frac{H_1 F_{\vec{r}}(t+s)}{\sqrt{ g(t)g(s)\,}\,} 
- \frac{2T_c}{\sqrt{g(t) g(s)\,}\,} \frac{H_1}{2T_c} \int_0^s \!\D\tau\, \delta(\tau) F_{\vec{r}}(t+s-2\tau) \nonumber \\
& & + 2T_c (ts)^{(2-d)/4} \int_0^s \!\D\tau\: \tau^{d/2-1} F_{\vec{r}}(t+s-2\tau)
\\
&\stackrel{s\to\infty}{\simeq}& s^{1-d/2} 2 T_c y^{(2-d)/4} \int_0^1 \!\D u\: u^{d/2-1} \left( 4\pi \left[ (y+1)-2u\right]\right)^{-d/2}
\exp\left[ - \frac{\vec{r}^2}{s} \frac{1}{y+1-2u}\right]
\nonumber
\EEA
where in the second line the singular contribution in $g(\tau)$ from (\ref{A3}) 
has been explicitly introduced to demonstrate the cancellation
of the $H_1$-dependent terms. In the third line, the regular large-time asymptotics of $g(t)$ 
has been inserted and finally, $F_{\vec{r}}(t)$ is expanded
for large times. For $d>2$, we now have $g(t)\sim t^0$ instead of $t^{d/2-1}$ and an analogous calculation gives
\BEQ 
C(t,s;\vec{r}) = s^{1-d/2}\, 2 T_c \int_0^1 \!\D u\: u^{d/2-1} \left( 4\pi \left[ (y+1)-2u\right]\right)^{-d/2}
\exp\left[ - \frac{\vec{r}^2}{s} \frac{1}{y+1-2u}\right]
\EEQ
which proves the assertion in section~4. For $\vec{r}=\vec{0}$, 
the two-time autocorrelator and its scaling function (\ref{4.10}) is readily obtained.  

Finally, we examine the special case $d=2$. Indeed, the treatment is analogous to the one of the 
$4D$ magnetic spherical model \cite{Hase06,Ebbi08}. From (\ref{A1}), one readily finds $\lap{g}(p)$. For large times, it can be shown that
$g(t)\simeq -H_1/(2 T_c) \delta(t) + \bar{g}_2/\ln t$ \cite{Hase06,Ebbi08,Fell71}. 
The spherical parameter $g(t)$ enters into the two-time response and correlation functions 
only through ratios $g(t)/g(s)\simeq \ln t/\ln s = \ln (ys)/\ln s\simeq 1 + \ln y/\ln s$. 
Therefore, these logarithmic terms only arise as {\em additive} logarithmic corrections to scaling, 
but do not affect the scaling form itself \cite{Hase06,Ebbi08}.
This can also be seen in the explicit results (\ref{4.7}),(\ref{4.10}), since the scaling functions are continuous in $d$, even at $d=2$.  
This is different, however, for the interface width, where the amplitudes eq. (\ref{4.11}) diverge for $d\to 2$. 
This case must therefore be analysed separately. In analogy with the treatment which led to (\ref{A4}) for $d<2$, we have first the 
cancellation of the $H_1$-dependent terms and then estimate the large-$t$ behaviour of the remaining integral 
\BEA
w^2(t) &\simeq& 2 T_c \ln t \int_{0}^{t} \!\D\tau\: \frac{1}{\ln\tau} F_{\vec{0}}(2t-2\tau) 
= 2T_c \ln t\: {\cal L}^{-1}\left( \left({\cal L}\left(\frac{1}{\ln \tau}\right)\right)(p) 
\left({\cal L} F_{\vec{0}}(2\tau)\right) (p) \right)(t)
\nonumber \\
&\simeq& \frac{2 T_c}{8\pi} \ln t\: \left({\cal L}^{-1} \frac{1}{p}\right)(t) \:\sim\: \ln t
\EEA
Again, the non-universal amplitude $\bar{g}_2$ cancels. 
We recognise the integral as a Laplace convolution and estimate the leading small-$p$ behaviour of the two factors. 
First, we recall $\left({\cal L}(1/\ln\tau)\right)(p)\sim -[p(C_E+\ln p)]^{-1}$, see \cite{Hase06,Ebbi08}. The second factor is  
the Laplace transform of $F_{\vec{0}}(2\tau)$ in $2D$, which is evaluated in analogy with similar integrals treated above, as follows 
(with a cut-off $\eta$; $E_1$ is the exponential integral \cite[eq. (5.1.11)]{Abra65})
\BD
\left({\cal L}F_{\vec{0}}(2\tau)\right)(p)=\int_0^{\infty} \!\D\tau\: e^{-(p+8)\tau} I_0^2(4\tau)) 
\stackrel{p\to 0}{\simeq} 
\int_{\eta}^{\infty} \!\D\tau\: e^{-p\tau} (8\pi \tau)^{-1} = \frac{1}{8\pi} E_1(p\eta)
\stackrel{p\to 0}{\simeq} -\frac{C_E + \ln p}{8\pi}
\ED
As in \cite{Ebbi08}, terms only depending on $\ln\eta$ must be absorbed into the `finite' contribution coming from the integral 
$\int_0^{\eta}$, which leads here to non-leading terms. 
This proves the middle line in (\ref{4.11}).

\subsection{Ageing at low temperatures}

For $T<T_c$, define the auxiliary variable $m^2 := 1-T/T_c$. First, we find $\lap{g}(p)$ from (\ref{4.1}), with $\lap{f}(p)$ explicitly
given in (\ref{A1}). This gives
\BEQ
\lap{g}(p) \simeq \left\{ \begin{array}{ll} 
\frac{H_1}{2T_c} \frac{1}{m^2} + \frac{H_1 A_d}{d m^4} p^{d/2} +\ldots & \mbox{\rm ~;~~ if $0<d<2$} \\
\frac{H_1}{2T_c} \frac{1}{m^2} + \frac{H_1}{m^4}{\lap{f}\,}'(0) p + \frac{H_1 A_d}{d m^4} p^{d/2} +\ldots & \mbox{\rm ~;~~ if $2<d<4$} 
\end{array} \right.
\EEQ
where we used $T_c + T/m^2 = T_c/m^2$. Further regular terms appear for $d>4,6,\ldots$, 
but will only give rise to corrections to the leading scaling behaviour. This gives in turn
\BEQ
g(t) \simeq \left\{ \begin{array}{ll}
\frac{H_1}{2T_c} \frac{1}{m^2}\, \delta(t) + \frac{H_1}{4 m^4}(8\pi)^{-d/2}\, t^{-1-d/2} + \ldots & \mbox{\rm ~;~~ if $0<d<2$} \\
\frac{H_1}{2T_c} \frac{1}{m^2}\, \delta(t) + \frac{H_1}{d m^4}{\lap{f}\,}'(0) \delta'(t) 
+ \frac{H_1}{4 m^4}(8\pi)^{-d/2}\, t^{-1-d/2} + \ldots & \mbox{\rm ~;~~ if $2<d<4$} 
\end{array} \right.
\EEQ

The response function is once more obtained by straightforward asymptotic expansion, 
which leads again to (\ref{4.6}) with the scaling function $f_R(y)$ being now given by (\ref{4.16}). 

Next, we analyse the interface width. The techniques to be used are quite close to the 
one applied in the critical case, but the leading term
turns out to be of a different form (let $d<2$ for simplicity)
\BEA
w^2(t) &=& \frac{H_1 F_{\vec{0}}(2t)}{g(t)} + 2 T \int_0^t \!\D\tau\: \frac{g(\tau)}{g(t)} F_{\vec{0}}(2t-2\tau) \nonumber \\
&\simeq&  \frac{H_1}{g(t)} \left( 1 + \frac{1}{m^2}\frac{T}{T_c}\right) F_{\vec{0}}(2t) 
+ 2T \int_0^t \!\D\tau\: \left(\frac{\tau}{t}\right)^{-1-d/2} F_{\vec{0}}(2t-2\tau)
\nonumber\\
&\simeq& 4 m^2 t + 2T t^{1+d/2} {\cal L}^{-1}\left( \left({\cal L}(\tau^{-1-d/2})(p)\right) \left({\cal L}F_{\vec{0}}(2\tau)(p)\right) \right)(t) 
\nonumber \\
&=& 4 m^2 t + 2T t^{1+d/2} \frac{\Gamma(-d/2)\Gamma(1-d/2)}{(8\pi)^{d/2}}\, \left({\cal L}^{-1} p^{d-1} \right)(t)
\nonumber \\
&=& 4 m^2 t + {\rm O}\left( T t^{1-d/2}\right)
\EEA
In the second line, we used both the singular and the leading regular term for $g(\tau)$. 
In contrast to the critical case, the singular contribution
does not cancel with the other $H_1$-dependent term, but combines into a new, 
leading term which depend on the ratio $T/T_c$ and vanishes when $T\to T_c$.
Therefore, the leading long-time behaviour only depends on the ratio $T/T_c$ and the short-ranged nature of the initial correlations. 
The `thermal' convolution term can be calculated via a Laplace convolution as before, but turns out to provide merely a correction to scaling. 
For $d>2$, the leading correction is O$(T)$ and the further derivatives of the $\delta$-function only generate, 
via partial integrations, further sub-leading corrections to scaling.  
Hence the leading result $w^2(t) \simeq 4 m^2 t$, valid for $t$ large enough, is independent of $d$, as asserted in (\ref{4.19}).  

The two-time autocorrelator is analysed in the same way (assume $d<2$ and let $y=t/s> 1$)
\typeout{*** saut de page *** } \newpage
\BEA
\lefteqn{C(t,s) = \frac{H_1 F_{\vec{0}}(t+s)}{\sqrt{g(t)g(s)\,}} 
+2T \int_0^s \!\D\tau\: \frac{g(\tau)}{\sqrt{ g(t)g(s)\,}}\, F_{\vec{0}}(t+s-2\tau)}
\nonumber \\
&\simeq& \frac{m^4 \Gamma(-d/2)d}{A_d (4\pi)^{d/2}} \frac{(ts)^{(2+d)/4}}{(t+s)^{d/2}}\left( 1 + \frac{1}{m^2}\frac{T}{T_c}\right) 
+\frac{2T}{(4\pi)^{d/2}} (ts)^{(2+d)/4} \int_0^s \!\D\tau\: \tau^{-1-d/2} (t+s-2\tau)^{-d/2} 
\nonumber \\
&=& 2^{d+2} m^2\: s\, y^{(2+d)/4} (y+1)^{-d/2} + {\rm O}\left( T s^{1-d/2}\right)
\EEA
as stated in (\ref{4.17}). For $d>2$, the leading correction is O$(T)$. 

Since all results are continuous in $d$ at $d=2$, a separate analysis of the $2D$ case is not necessary. 

\appsection{B}{On the Yeung-Rao-Desai inequalities}

The analogue of the well-known YRD-inequality \cite{Yeun96}, 
originally formulated for ageing magnetic systems, is: {\it for a growing interface with
a non-conserved dynamics and an uncorrelated initial state, the autocorrelation exponent $\lambda_C$ satisfies the bound}
\BEQ \label{B1}
\lambda_C \geq \demi \left( d + z b\right)
\EEQ
Such bounds may serve as checks on exponents, estimated from simulational or experimental data 
(i.e. for phase-ordering magnets with $T<T_c$, one has $b=0$, hence $\lambda_C\geq d/2$ \cite{Yeun96}). 

To see this, recall first the argument \cite{Yeun96} to bound the two-time height autocorrelator
\BEQ \label{B2}
C(t,s) = \int\!\D\vec{k}\: \wht{C}(t,s;\vec{k},-\vec{k}) \leq \int\!\D\vec{k}\: \sqrt{ \wht{C}(t,\vec{k})\, \wht{C}(s,\vec{k})\,} 
\EEQ
Eq. (1.2) gives the scaling form of the single-time correlator $\wht{C}(t,\vec{k})=\wht{C}(t,t;\vec{k},-\vec{k})$
\BD
\wht{C}(t,\vec{k}) = \frac{1}{\sqrt{V}} \int_{V} \!\D\vec{r}\: e^{-\II \vec{k}\cdot\vec{r}}\, t^{-b} F_C\left(1; \vec{r} t^{-1/z}\right) 
= L(t)^{d-zb}\, {\cal C}\left(k L(t)\right)
\ED
where the typical length scale $L(t)\sim t^{1/z}$ and ${\cal C}$ is a scaling function. 
In order to estimate the integral, recall that
the height fluctuations $\delta\wht{h}(t,\vec{k})$ and $\delta\wht{h}(s,-\vec{k})$ in 
$\wht{C}(t,s;\vec{k},-\vec{k})$ should become uncorrelated 
over distances larger than $\Delta L \gtrsim 2\pi {\tt a}/|\vec{k}|$ such that 
$\wht{C}(t,s;\vec{k},-\vec{k})\to 0$ rapidly if 
$\left(L(t)-L(s)\right)|\vec{k}|\gg 1$ \cite{Yeun96}. For uncorrelated initial conditions, 
one has $\lim_{s\to 0} \wht{C}(s,\vec{k}) \sim \lim_{\vec{k}\to\vec{0}} \wht{C}(s,\vec{k})= {\rm O}(1)$. 
In the scaling limit with $y=t/s$ large enough, the bound (\ref{B2}) gives
\BEA
\lefteqn{ \lim_{t,s\to\infty} C(t,s) \sim L(t)^{-\lambda_C} } \nonumber \\
&\leq& L(t)^{(d-zb)/2}\, C_d \int_0^{2\pi{\tt a}/L(t)} \!\!\D k\: k^{d-1}\, {\cal C}^{1/2}\left( k L(t)\right) 
\:=\: L(t)^{-(d+zb)/2}\: C_d \int_0^{2\pi{\tt a}} \!\D u\: u^{d-1}\, {\cal C}^{1/2}(u) \nonumber
\EEA
($C_d$ is a constant) and (\ref{B1}) follows. 

Alternatively, if the waiting time $s$ is itself already in the scaling regime, 
one has $\wht{C}(s,\vec{k})\stackrel{|\vec{k}|\to 0}{\sim} k^{zb-d}$
which leads to $\lambda_C\geq zb$, with an magnetic analogue already given in \cite{Yeun96}. 

\noindent 
{\bf Acknowledgements:}  This work was done during the workshop ``Advances in Non-equilibrium Statistical Mechanics''. 
We gratefully thank the organisers and the
Galileo Galilei Institute for Theoretical Physics for their warm and generous hospitality and the INFN for partial support.  
We thank N. Allegra, J.-Y. Fortin, H. Park, A. Pikovsky and U.C. T\"auber for useful discussions. This work 
was also partly supported by the Coll\`ege Doctoral franco-allemand Nancy-Leipzig-Coventry
({\it `Syst\`emes complexes \`a l'\'equilibre et hors \'equilibre'}) of UFA-DFH 
and also by Mid-career Researcher Program through NRF grant No.~2010-0026627 funded by the MEST. 


\end{document}